\documentclass[11pt]{article}
\usepackage{geometry}                % See geometry.pdf to learn the layout options. There are lots.
\usepackage{natbib}
\usepackage{amsthm}
\usepackage{amsmath}
\usepackage{natbib}
\usepackage[colorlinks,citecolor=blue,urlcolor=blue, filecolor=blue,backref=page]{hyperref}
\usepackage{graphicx}

\usepackage{amsthm,amsmath}
\RequirePackage[OT1]{fontenc}
\RequirePackage{amsthm,amsmath}
\newtheorem{definition}{Definition}

\newcounter{lastnote}

\usepackage{amsthm,amsmath}
\RequirePackage[OT1]{fontenc}
\RequirePackage{amsthm,amsmath}

\RequirePackage[colorlinks,citecolor=blue,urlcolor=blue]{hyperref}
\usepackage[latin1] {inputenc}
\usepackage{vmargin}
\usepackage{multirow}
\usepackage{subfigure}
\usepackage[normalem]{ulem}	
\usepackage{caption}

\usepackage{tikz} 
\usetikzlibrary{arrows,decorations.markings,chains,shapes.arrows,shapes.misc,fit, shapes}

 \tikzset{
  big arrow/.style={
    decoration={markings,mark=at position 1 with {\arrow[scale=2,#1]{>}}},
    postaction={decorate},
    shorten >=0.4pt},
  big arrow/.default=black}
  
    \tikzset{
  dash arrow/.style={
    decoration={markings,mark=at position 1 with {\arrow[scale=1.2,#1]{>}}},
    postaction={decorate},
    shorten >=0.4pt},
  dash arrow/.default=black}
  
  \tikzset{ doppia/.style={
  color=black, cylinder, draw, shape border rotate=90, aspect=0.15, text width=5.5em, minimum height=6.5em, minimum width=2em, align=center}}

\theoremstyle{plain}

\newtheorem{lemma}{Lemma}[section]
\newtheorem{corollary}{Corollary}[section]

\usepackage{amssymb}
\usepackage{epsfig}

\DeclareMathAlphabet{\mathpzc}{OT1}{pzc}{m}{it}

\newcommand{\LR}{\textrm{LR}}

\usepackage[makeroom]{cancel}
\begin{document}

%% *** Frontmatter *** 

\title{A nonparametric Bayesian approach to the rare type match problem}

%\thankstext{T1}{<thanks text>}

\author
{Giulia Cereda,\\
%\normalsize{An Unknown Address, Wherever, ST 00000, USA}\\
\normalsize{Leiden University, Mathematical Institute}\\
Richard Gill,\\
%\normalsize{An Unknown Address, Wherever, ST 00000, USA}\\
\normalsize{Leiden University, Mathematical Institute}\\
\normalsize{Combray Causality Consulting}}
\maketitle
\begin{abstract}
The ``rare type match problem'' is the situation in which, in a criminal case, the suspect's DNA profile, matching the DNA profile of the crime stain, is not in the database of reference.
Ideally, the evaluation of this observed match in the light of the two competing hypotheses (the crime stain has been left by the suspect or by another person) should be based on the calculation of the likelihood ratio and depends on the population proportions of the DNA profiles, that are unknown. 
We propose a Bayesian nonparametric method that uses a two-parameter Poisson Dirichlet distribution as a prior over the ranked population proportions, and discards the information about the names of the different DNA profiles.
This model is validated using data coming from European Y-STR DNA profiles,
and the calculation of the likelihood ratio becomes quite simple thanks to an Empirical Bayes approach for which we provided a motivation. 
\end{abstract}

%% ** Keywords **

Forensic statistics, likelihood ratio, rare type match, Bayesian nonparametric

\section{Introduction}

The largely accepted method for evaluating how much some available data $\mathcal{D}$ (typically forensic evidence) helps discriminate between two hypotheses of interest (the prosecution hypothesis $H_p$ and the defense hypothesis $H_d$), is the calculation of the \emph{likelihood ratio} (LR), a statistic that expresses the relative plausibility of the data under these hypotheses, defined as
\begin{equation}
\label{eqa}
\LR=\frac{\Pr(\mathcal{D}|H_p)}{\Pr(\mathcal{D}|H_d)}.
\end{equation}

Widely considered the most appropriate framework to report a measure of the `probative value' of the evidence regarding the two hypotheses \citep{robertson:1995, evett:1998, aitken:2004,balding:2005}, it indicates the extent to which \textcolor{black}{observed data support one hypothesis over the other. The likelihood ratio is supposed to be multiplied to the prior odds, in order to obtain the posterior odds. The latter is the quantity of interest for a judge, but the prior odds do not fall within the statistician competence. Even if a judge does not explicitly do the Bayesian updating, the likelihood ratio is still considered to be the correct way for the expert to communicate their evaluation of the weight of the evidence to the court. We refer the reader to \citet{taroni:2006} for an extensive dissertation on the use of  likelihood ratios in forensic statistics.} 
Forensic literature presents many approaches to calculate the LR, mostly divided into Bayesian and frequentist methods (see \citet{cereda:2015, cereda:2015b} for a careful differentiation between these two approaches). 

This paper proposes a first application of a Bayesian nonparametric method to assess the likelihood ratio in the rare type match case, 
the challenging situation in which there is a match between some characteristic of the recovered material and of the control material, but this characteristic has not been observed before in previously collected samples (i.e.\ in the database of reference). 
This constitutes a problem because the value of the likelihood ratio depends on the unknown proportion of the matching characteristic in a reference population, and the uncertainty over this proportion, in standard practice for simpler situations, is dealt with using the relative frequency of the characteristic in the available database. 
In particular, we will focus on Y-STR data, for which the rare type match problem keeps turning up \citep{cereda:2015b}. The  problem is so substantial that it has been called ``the fundamental problem of forensic mathematics'' \citep{brenner:2010}.
%In this case data to evaluate is made of the information about the matching profile and of the list of DNA profiles in the database.

The use of our Bayesian nonparametric method involves the mathematical assumption that there are infinitely many different Y-STR profiles. Of course, we do not believe this literally to be true. We do suppose that there are so many profiles that we cannot say anything sensible about their exact number, except that it is very large. Hence, we pretend they are infinitely many, so that we can use the chosen Bayesian nonparametric method. 
The parameter of the model is the infinite-dimensional vector $\mathbf{p}$, containing the (unknown) sorted population proportions of all possible Y-STR profiles. As a prior over $\mathbf{p}$ we choose the two-parameter Poisson Dirichlet distribution, and we model the uncertainty over its parameters $\alpha$ and $\theta$ through the use of a hyperprior. 
The information contained in the \textcolor{black}{ actual numbers, a list of which form the name of each} Y-STR profiles is discarded, thereby reducing the full data $\mathcal{D}$ to a smaller set $D$.

\textcolor{black}{If compared to traditional Bayesian methods such as those discussed in \citet{cereda:2015}, this method has the advantage of having a prior for the parameter \textbf{p} that is more realistic for the population we intend to model.
Moreover, despite its technical theoretical background, we empirically derived an approximation that makes the method \textcolor{black}{intuitive and simple} to apply for practical use: indeed, simulation experiments show that a hybrid empirical approach that plugs in maximum likelihood estimators for the hyperparamenter is justified, at least when using populations that look like the complete Y-STR data from European populations.
The last point in favour of the choice of the two-parameter Poisson Dirichlet prior over \textbf{p} is that it has the following sufficiency property: the probability of observing a new Y-STR profile only depends on the number of already observed Y-STR profiles and on the sample size, while the probability of observing a Y-STR profile that is already in the database only depends on its frequency in the database and on the sample size.
}

%The reduction of the data can be a wise practice in presence of many nuisance parameters as explained in \citet{cereda:2015b}, and  sometimes the likelihood ratio based on the data reduction is much more precisely estimated than the likelihood ratio based on all data.

The paper is structured as follows: Section~\ref{rare} discusses the state of the art regarding the rare type match problem and the evaluation of Y-STR matches. Section~\ref{m} presents our model, with the assumptions and the prior distribution chosen for the parameter $\mathbf{p}$ along with some theory on random partitions and the Chinese restaurant representation, useful to provide a prediction rule and a \textcolor{black}{convenient and compact representation of the reduced data $D$}. Also, a lemma that facilitates computing the likelihood ratio in a very simple way is presented and proved.
In Section~\ref{lr}, the likelihood ratio is derived.
%An alternative representation of the same model via the two-parameter Chinese restaurant process is also described.%which can be used for all the situations in which prosecution and defense agree on the distribution of part of the data and disagree on the distribution of the rest, given the parameter(s). This result will allow to derive the LR in a very elegant way (Section~\ref{lr}).
Section~\ref{ard} illustrates the application of this model to a database sampled from an artificial population. We will discuss data-driven choices for the hyperparameters, and the derivation of the likelihood ratio values obtained both with and without reducing the data to partitions, in the ideal situation in which vector $\mathbf{p}$ is known. Also, the distribution of the likelihood ratios for different rare type match cases is analysed, along with the analysis of two different errors.

\section{State of the art}\label{rare}
\textcolor{black}{Y-STR data have been our main motivation for studying the rare type match problem. Our model will 
reduce the relationship among various profiles to a binary ``match'' or ``no match'' equivalence relation.
However, there is a big debate in the scientific community regarding whether it is acceptable to throw away the genetic structure of this kind of data.
In this section we discuss the state of the art regarding the rare type match problem as a general issue and also the state of the art regarding methods to assess evidential values of matching  Y-STR profiles. Indeed, the rare type match problem is an interesting problem also outside the Y-STR profile setting and our model can be applied also to other kinds of data.}
\subsection{The rare type match problem}

The evaluation of a match between the profile of a particular piece of evidence and a suspect's profile depends on the relative frequencies of that profile in the population of potential perpetrators. 
Indeed, it is intuitive that the rarer the matching profile, the more the suspect is in trouble. 
Problems arise when the observed frequency of the profile in a sample (database) from the population of interest is 0. This problem can be named as ``the new type match problem", but we decided to use the name ``rare type match problem", motivated by the fact that a profile that has zero occurrences is likely to be rare, even though it is challenging to quantify how rare it is. The rare type match problem is particularly important for new kinds of forensic evidence, \textcolor{black}{such as results from DIP-STR markers} (see for instance \citet{cereda:2014b}) for which the available database size is still limited. 
The problem also occurs when more established types of evidence, such as Y-chromosome
(or mitochondrial) DNA profiles are used, \textcolor{black}{as explained in Section \ref{y}}, \textcolor{black}{and they have been our  main motivation for the present study}.

The rare type match problem has been addressed in well known non-forensic statistics domains, and many solutions have been proposed. The \emph{empirical frequency estimator}, also called \emph{naive estimator}, that uses the frequency of the characteristic in the database, puts unit probability mass on the set of already observed characteristics, and it is thus unprepared for the observation of a new type. 
A solution could be the \emph{add-constant} estimators (in particular the well-known \emph{add-one} estimator, due to \citet{laplace:1814}, and the \emph{add-half} estimator of \citet{krichevsky:1981}), which add a constant to the count of each type, included the unseen ones. However, these methods require knowledge of the number of possible unseen types, and they perform badly when this number is large compared to the sample size (see \citet{gale:1994} for an additional discussion). Alternatively, \citet{good:1953}, based on an intuition on A.M. Turing, proposed the \emph{Good-Turing estimator} for the total unobserved probability mass, based on the proportion of singleton observations in the sample. An  \textcolor{black}{application} of this estimator to the frequentist LR assessment in the rare type match case is proposed in \citet{cereda:2015b}. 

%For a comparison between the \emph{add one} and the \emph{Good-Turing} estimator, see \citet{orlitsky:2003}.
%As pointed out in \citet{anevski:2013}, the \emph{naive estimator}, and the \emph{Good Turing estimator} are in some sense complementary: the first gives a good estimate for the observed types, and the second for the probability mass of the unobserved ones. 
Recently, \citet{orlitsky:2004} have introduced the \emph{high-profile estimator}, which extends the tail of the \emph{naive estimator} to the region of unobserved types.  \citet{anevski:2013} improved this estimator and provided a consistency proof for their modified estimator, while original authors only provided heuristic reasoning that turned out to be rather difficult to make rigorous.

Bayesian nonparametric estimators for the probability of observing a new type have been proposed by \citet{tiwari:1989} using Dirichlet processes, by \citet{lijoi:2007, deblasi:2015} using a general Gibbs prior, and by \citet{favaro:2009} with specific focus on the two-parameter Poisson Dirichlet prior, for which \citet{arbel:2015} provides large sample asymptotic and credible bands. In particular, \citet{favaro:2014} shows the link between the Bayesian nonparametric approach and the Good-Turing estimator. 
However, the LR assessment requires not only the probability of observing a new species but also the probability of observing this same species twice (according to the defence, the crime stain profile and the suspect profile are two independent observations): to our knowledge, the present paper is the first to address the problem of assessment of the LR in the rare type match case using a Bayesian nonparametric model. As a prior for $\mathbf{p}$ we will use the two-parameter Poisson Dirichlet distribution, which is proving useful in many discrete domains, in particular language modeling \citep{teh:2006}. Besides, it predicts a power-law behavior that describes an incredible variety of phenomena \citep{newman:2005}, among which the distribution of Y-STR haplotypes, too.

\subsection{Evaluation of matching probabilities of Y-STR data}\label{y}

 Y-STR profiles are typically used to detect male DNA in male-female DNA mixtures and
 are made of a number (usually varying from 7 to 23) of STR  polymorphisms belonging to the non-recombining part of the Y-chromosome. Hence, there is no biological reason to assume independence among Y-STR loci.   
Even though the lack of recombination is in principle balanced by recurrent and backward mutations, the existence of such a dependency is studied and confirmed by \citet{caliebe:2015}. 
For what concerns Y-STR population frequencies, the dependency between loci implies that no factorization (of the kind used for the autosomal markers) can be used to calculate these frequencies, and that the available databases are too small with respect to the large space of possible profiles, hence containing a high proportion of singletons. Indeed, the rare type match case is very frequent when using Y-STR data, and the use of simplistic methods such as the profile count is too conservative for practical use (it is bounded from below by the inverse of the
database size) \citep{caliebe:2015}.
In \citet{andersen:2018} and \citet{andersen:2019}, approximations of the joint distribution with second and third order dependencies between loci are explored. However, as admitted by the authors, there is a limitation due to the inadequacy of the sizes of available databases  that makes it necessary to use  simulations, that in turns are oversimplification of real data.

Moreover, as highlighted already in 1994 by \citet{balding:1994} match probabilities cannot be identified with population frequencies since a match can be due also to a certain degree of relatedness between the two donors of the stain. This is particularly true for Y-STR data, since Y-STR profiles are inherited almost identical from father to son. More recently, \citet{andersen:2017} investigate the influence of relatedness on matches and make a study concluding that 95$\%$ of matching profiles are separated by a relatively small number (50-100) of meiosis, hence the degree of relatedness is a very influential factor, according to their study. They thus propose a method to describe the distribution of the number of males with a matching Y-STR profile, extending the approach to mixtures in \citet{andersen:2019b}. One limitation of this study is that it is based on extensive simulations which have to be performed anew in each new application,  on   assumptions about genetic evolutionary model, and on parameters hard to be validated.

\textcolor{black}{There is a huge number of methods developed to assess the evidential values for Y-STR data. 
Among those that are developed precisely for the rare type match case there are \citet{egeland:2008}, \citet{brenner:2010}, \citet{cereda:2015b} and \citet{cereda:2015}.  
All these methods do not take into account genetic information contained in the allelic numbers forming a Y-STR DNA profile. For instance, due to relatedness, the observation of a particular Y-STR profile increases the probability of observing the same Y-STR profile again or Y-STR profiles that differ only for few alleles. We refer the reader to \citet{Roewer:2009, Buckleton:2011, willuweit:2011, Wilson:2003} for models that use population genetics for coancestry. These models are not designed to be used for the rare type match case, but the Discrete Laplace method presented in \citet{andersen:2013b} can be successfully applied to that purpose, as shown in \citet{cereda:2015b}.}

\textcolor{black}{
After a careful study of the available methods for assessing likelihood ratios (or matching probabilities) for Y-STR matches, one can see that they are of different nature (some of them do their best to exploit  the genetic structure, others don't) and based on different assumptions.  In our opinion none of them is fully satisfactory  and at the same time useful for the rare type match and for general cases. 
As far as we are concerned, we decided to asses the probability of the reduction of the data taking into account only the equalities and inequalities among profiles  rather than considering the specific Y-STR observed characteristics. We know part of the scientific community will not agree with our approach, precisely because of the results shown in \citet{andersen:2017}, but we believe in the accuracy of our method.
Moreover, even though  Y-STR data have been the main motivation for  this study,  this model  is actually applicable to different kinds of data (in 
principle for all forensic data that shows power law behaviour). When applied to data without genetic structures (such as tire marks or glass fragments), these kind of criticisms should fade away.  }

The Y-STR marker system will thus be employed here as an extreme but in practice common and important way in which the problem of assessing the evidential value of rare type match can arise, but we don't pretend to solve the problem entirely. \textcolor{black}{ We believe that the analyst should perform several analyses using different models and different assumptions, and compare the performance of the different methods, in order to  try to learn from the differences (or lack of differences) between the conclusions which would follow from each method individually. }

\textcolor{black}{The big issues of working with Y-STR data is the unavailability of reliable databases, which are representative of actual population. The YHRD database is in fact a collection of databases coming from police or laboratories. We are well aware of this limitation.}
 \section{The model}\label{m}

 \subsection{Notation and data}
Throughout the paper the following notation is chosen: random variables and their values are denoted, respectively, with uppercase and lowercase characters: $x$ is a realization of $X$. Random vectors and their values are denoted, respectively, by uppercase and lowercase bold characters: $\mathbf{p}$ is a realization of the random vector $\mathbf{P}$. Probability is denoted with $\Pr(\cdot)$, while the density of a continuous random variable $X$ is denoted alternatively by $p_{X}(x)$ or by $p(x)$ when the subscript is clear from the context. For a discrete random variable $Y$, the density notation $p_Y(y)$ and the discrete one $\Pr(Y=y)$ will be interchangeably used.
Moreover, we will use shorthand notation like $p(y \mid x)$ to stand for the probability density of Y with respect to the conditional distribution of $Y$ given $X = x$.

Notice that in Formula~\eqref{eqa}, $\mathcal{D}$ was regarded as the event corresponding to the observation of the available data. However, later in the paper, $\mathcal{D}$ will be regarded as a random variable generically representing the data. The particular data at hand will correspond to the value $d$. In that case, the following notation will thus be preferred:
\begin{equation*}
\LR=\frac{\Pr(\mathcal{D}=d|H=h_p)}{\Pr(\mathcal{D}=d|H=h_d)}\quad \text{or} \quad \frac{p(d|h_p)}{p(d|h_d)}.
\end{equation*}

Lastly, notice that ``DNA types'' is used throughout the paper as a general term to indicate Y-STR profiles.

\textcolor{black}{The data used in the present study were obtained from the
Y Chromosome Haplotype Reference Database (YHRD) \citep{willuweit:2007, purps:2014}. Here, only 7 of the markers included
in the PowerPlex1Y23 system (PPY23, Promega Corporation, Madison, WI) were investigated : DYS19, DYS389I, DYS389II.I, DYS390, DYS391, DYS392, DYS393. The dependence between these 7 ``core markers'' is studied in \citet{caliebe:2015} that concludes that ``each of these seven markers contribute indispensable information about each other markers from the same set''. 
}

\subsection{Model assumptions}\label{m-a}

Our model is based on the two following assumptions:

\begin{description}
\item[Assumption 1] There are infinitely many different DNA types in Nature.
\end{description}
    \textcolor{black}{This assumption, already used by e.g.\ \citet{kimura:1964} in the `infinite alleles model', allows the use of Bayesian nonparametric methods and is very useful for instance in `species sampling problems' when the total number of possible different species in Nature cannot be specified. This assumption is sensible also in case of Y-STR DNA profiles since the state space of possible different haplotypes is so large that it can be considered infinite.}
\begin{description}

\item[Assumption 2] The names of the different DNA types do not contain relevant information.

\end{description}
Actually, the specific sequence of numbers that forms a DNA profile carries information: if two profiles show few differences this means that they are separated by few mutation drifts, hence the profiles share a relatively recent common ancestor. However, this information can be very difficult to use and \textcolor{black}{it might be wiser not to try to use it in the LR assessment.}
This is the reason why we will treat DNA types as ``colours'', and only consider the partition into different categories. Stated otherwise, we put no topological structure on the space of the DNA types. 

Notice that this assumption makes the model a priori suitable for any characteristic which has many different possible types \textcolor{black}{showing power law behaviour}, thus the approach described in this paper still holds, in principle, after replacing `DNA types' with any other category. 

\subsection{Prior}\label{prior}

In Bayesian statistics, parameters of interest are modeled through random variables.
The (prior) distribution over a parameter should represent the prior uncertainty about its value.

The assessment of the LR for the rare type match involves two unknown parameters of interest: one is $h \in \{h_p, h_d\}$, representing the unknown true hypothesis, the other is $\mathbf{p}$, the vector of the unknown population frequencies of all DNA profiles in the population of potential perpetrators.
The dichotomous random variable $H$ is used to model parameter $h$, and the posterior distribution of this random variable, given the data, is the ultimate aim of the forensic inquiry.
Similarly, a random variable $\mathbf{P}$ is used to model the uncertainty over $\mathbf{p}$. 
Because of Assumption~1, $\mathbf{p}$ is an infinite-dimensional parameter, hence the need for Bayesian nonparametric methods \citep{hjort:2010, ghosal:2015}.
In particular, because of Assumption~2, data can be reduced to partitions, as explained in Section~\ref{partitions}, and it will turn out that the distribution of these partitions does not depend on the order of the $p_i$. 
Hence, we can define the parameter $\mathbf{p}$ as having values in $\nabla_{\infty}=\{(p_1, p_2, ...) \mid p_1\geq p_2 \geq ..., \sum p_i=1, p_i>0\}$, the ordered infinite-dimensional simplex.
The uncertainty about its value will be expressed by the two-parameter Poisson Dirichlet prior \citep{pitman:1997,  feng:2010, buntine:2012, pitman:2006}. 

The two-parameter Poisson-Dirichlet distribution can be defined through the following stick-breaking representation \citep{ghosal:2015}: 

\begin{definition}[two-parameter GEM distribution]
Given $\alpha$ and $\theta$ satisfying the following conditions:
\begin{equation}\label{condition}
0\leq \alpha<1, \text{ and } \theta>-\alpha.
\end{equation}
the vector $\mathbf{W}=(W_1,W_2,...)$ is said to be distributed according to the \emph{GEM($\alpha, \theta$)}, if $$\forall i\quad W_i=V_i\prod_{j=1}^{i-1} (1-V_j),$$ where $V_1$, $V_2$,... are independent random variables distributed according to $$V_i\sim \text{Beta}(1-\alpha, \theta+ i\alpha).$$ 
It holds that $W_i >0$, and $\sum_{i}W_i=1.$ 
\end{definition}
The GEM distribution (short for Griffin-Engen-McCloskey distribution') is well-known in the literature as the ``stick-breaking prior'' since it measures the random sizes in which a stick is broken iteratively. 
%This distribution is invariant under the so-called size-biased permutations}\citep{engen:1975}, \sout{that is the random permutation defined by sampling from the population and assigning to each type a label, based on the order in which the types are first sampled.}

\begin{definition}[Two-parameter Poisson Dirichlet distribution]\label{pd}
Given $\alpha$ and $\theta$ satisfying condition~\eqref{condition}, and a vector $\mathbf{W}=(W_1,W_2,...) \sim \text{GEM}(\alpha, \theta)$, the random vector $\mathbf{P}=(P_1, P_2, ...)$ obtained by ranking $\mathbf{W}$, such that $P_i\geq P_{i+1}$, is said to be \emph{Poisson Dirichlet distributed} PD$(\alpha, \theta)$. Parameter $\alpha$ is called the \emph{discount parameter}, while $\theta$ is the \emph{concentration parameter.}

\end{definition}

For our model \textcolor{black}{we will not allow $\alpha=0$}, hence we will assume $0<\alpha<1$, in order to have a prior that shows a power-law behavior as the one observed in the YHRD database (see Section~\ref{fit}).

\textcolor{black}{The main reason that prompted us to choose the two-parameter Poisson Dirichlet distribution among the possible Bayesian nonparametric priors is given by model fitting (see Section~\ref{fit}). However, there is a very nice feature of this model. It is the only one that has the following very convenient sufficiency property \citep{zabell:2005}: the probability of observing a new species only depends on the number of already observed species and on the sample size, and the probability of observing an already seen species only depends on its frequency in the sample and on the sample size.
}

\textcolor{black}{Lastly, we point out that, in practice, we cannot assume that we know the parameters $\alpha$ and $\theta$: we will resolve this by using a hyperprior. 
}
\subsection{Bayesian network representation of the model}
The typical data to evaluate in case of a match is $\mathcal{D}=(E, B)$, where $E=(E_s, E_t)$, and\\

\hfill\begin{minipage}{\dimexpr\textwidth-0.5cm}
 $E_s$ = suspect's DNA type,\\
 $E_t$ = crime stain's DNA type (matching the suspect's type),\\
 $B$ = a reference database of size $n$, which is treated here as a random sample of DNA types from the population of possible perpetrators.\\
\end{minipage}

\noindent The hypotheses of interest for the case are:\\

\begin{figure}[htbp]

\centering
\begin{tikzpicture} [scale=0.85]
\node[draw, ellipse, minimum width=1.2cm]                (t) at (-4,4)  { $A, \Theta$};
\node [draw,ellipse, minimum width=1.2cm]              (h) at (3,2) { $H$};
\node [draw,ellipse, minimum width=1.2cm]              (p) at (-4,2) { $\mathbf{P}$};
\node [draw, ellipse, minimum width=1.2cm]              (y1) at (-8, -1) { $X_1$};
\node [draw, ellipse, minimum width=1.2cm]              (y2) at (-6,-1) { $X_2$};
\node [ ellipse, minimum width=1.2cm]              (y24) at (-3.5,-1) { $...$};
\node [draw,ellipse, minimum width=1.2cm]              (yn) at (-1,-1) { $X_n$};
\node [draw,ellipse, minimum width=1.2cm]              (yn33) at (1,-1) { $X_{n+1}$};
\node [draw, ellipse, minimum width=1.2cm]              (yn34) at (4,-1) { $X_{n+2}$};
\node [draw, ellipse, minimum width=1.2cm]              (d) at (-3,-3) { $D$};

   \draw[black, big arrow] (t) -- (p);
   \draw[black, big arrow] (p) -- (y1); 
   \draw[black, big arrow] (p) -- (y2); 
   \draw[black, big arrow] (p) -- (yn); 
      \draw[black, big arrow] (p) -- (yn33); 
    \draw[black, big arrow] (p) -- (yn34); 
    \draw[black, big arrow] (h) -- (yn34); 
    \draw[black, big arrow] (yn33) -- (yn34); 
  \draw[black, big arrow] (y1) -- (d); 
    \draw[black, big arrow] (y2) -- (d); 
      \draw[black, big arrow] (yn) -- (d); 
        \draw[black, big arrow] (yn33) -- (d); 
  \draw[black, big arrow] (yn34) -- (d);

\end{tikzpicture}
\caption{\small Bayesian network showing the conditional dependencies of the relevant random variables in our model.}
\label{berhnt}

\end{figure}
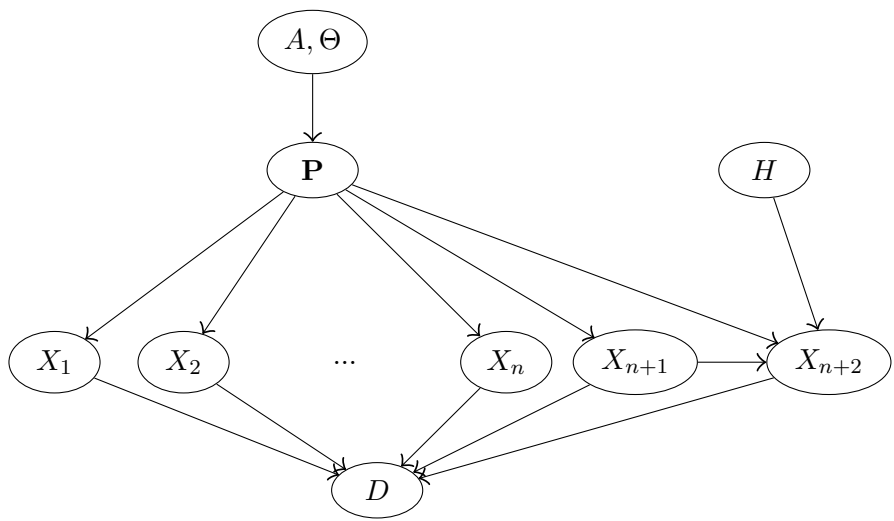

\hfill\begin{minipage}{\dimexpr\textwidth-0.5cm}

 $h_p$ = The crime stain originated from  the suspect,\\
 $h_d$ = The crime  stain originated from  someone else.\\
\end{minipage}
In agreement with Assumption~2, the model will ignore information about the names of the DNA types: data $\mathcal{D}=(E, B)$ will thus be reduced to $D$ accordingly.
The Bayesian network of Figure~\ref{berhnt} encapsulates the conditional dependencies of the random variables $(H, A, \Theta, \mathbf{P}, X_1, ..., X_{n+2}, D)$, whose joint distribution is defined below in terms of the conditional distribution, using the factorization implied by the Bayesian network itself.

$H$ is a dichotomous random variable that represents the hypotheses of interest and can take values $h \in \{h_p, h_d\}$, according to the prosecution or the defense, respectively. A uniform prior on the hypotheses is chosen \textcolor{black}{for mathematical convenience since it will not affect the likelihood ratio (the variable $H$ being in the conditioning part)}:
$$\Pr(H=h)\propto 1 \quad \textrm{for} \ h \in \{h_p, h_d\}.$$

 ($A, \Theta$) is the random vector that represents the hyperparameters $\alpha$ and $ \theta$, satisfying condition~\eqref{condition}. The joint prior density of these two parameters will be generically denoted as $p(\alpha, \theta)$:
$$(A, \Theta) \sim p(\alpha, \theta).$$ For obvious reasons, this will be called the `hyperprior' throughout the text.

The random vector $\mathbf{P}$ with values in $\nabla_{\infty}$ represents the ranked population frequencies of Y-STR profiles. $\mathbf{P}=(p_1, p_2, ...)$ means that $p_1$ is the frequency of the most common DNA type in the population, $p_2$ is the frequency of the second most common DNA type, and so on. As a prior for $\mathbf{P}$ we use the two-parameter Poisson Dirichlet distribution:
$$\mathbf{P}| A=\alpha,  \Theta= \theta \sim PD(\alpha, \theta).$$

The database is assumed to be a random sample from the population. Integer-valued random variables $X_1$, ..., $X_n$ are here used to represent the (unknown) ranks in the population of the frequencies of the DNA types in the database. For instance, $X_3 = 5$ means that the third individual in the database has the fifth most common DNA type in the population.
%Since $\mathbf{p}$ is unknown these random variables cannot be observed. 
Given $\mathbf{p}$ they are an i.i.d. sample from $\mathbf{p}$:
\begin{equation} \label{eqx}
X_1, X_2, ..., X_n | \mathbf{P}=\mathbf{p} \sim_{i.i.d.} \mathbf{p}.
\end{equation}

 $X_{n+1}$ represents the rank in the population ordering of the suspect's DNA type. It is again an independent draw from $\mathbf{p}$.
$$ X_{n+1} | \mathbf{P}=\mathbf{p} \sim  \mathbf{p}.$$

$X_{n+2}$ represents the rank in the population ordering, of the crime stain's DNA type. According to the prosecution, given $X_{n+1}=x_{n+1}$, this random variable is deterministic (it is equal to $x_{n+1}$ with probability 1). According to the defence, it is another sample from $\mathbf{p}$, independent of the previous ones:
\begin{equation}\label{sdf} 
X_{n+2} |  \mathbf{P}=\mathbf{p} , X_{n+1}=x_{n+1}, H=h \sim 
\begin{cases} 
\delta_{x_{n+1}} & \text{if } h=h_p\\
\mathbf{p} &  \text{if } h=h_d
\end{cases}.
\end{equation}

In order to observe $X_1$, ..., $X_{n+2}$, one would need, by definition, to know the rank, in terms of population proportions, of the frequency of each DNA type in the database. This is not known, hence $X_1, ..., X_n$ are not observed.

Section~\ref{partitions} recalls some notions about random partitions, useful before defining node $D$, the `reduced' data that we want to evaluate.

\subsection{Random partitions and database partitions}\label{partitions}

A \emph{partition of a set $S$} is an unordered collection of nonempty and disjoint subsets of $S$, the union of which forms $S$. 
Particularly interesting for our model are partitions of the set $S=[n]=\{1, ..., n\}$, denoted as $\pi_{[n]}$. The set of all partitions of $[n]$ will be denoted as $\mathcal{P}_{[n]}$. Random partitions of $[n]$ will be denoted as $\Pi_{[n]}$, $n\in \mathbb{N}$. Also, a \emph{partition of $n$} is a finite nonincreasing sequence of positive integers that sum up to $n$. Partitions of $n$ will be denoted as $\pi_n$, while random partitions as $\Pi_n$.

%
%A \emph{partition of $n$} is a finite non increasing sequence of positive integers that sum up to $n$. Partitions of $n$ will be denoted as $\pi_n$. The set of all partitions of $n$ will be denoted as $\mathcal{P}_{n}$. Random partitions of $n$ will be denoted as $\Pi_{n}$.
%

Given a sequence of integer valued random variables $X_1, ..., X_n$, let $\Pi_{[n]}(X_1, X_2, ..., X_n)$ be the random partition defined by the equivalence classes of their indices using the random equivalence relation $i \sim j$ if and only if $X_i=X_j$.
This construction allows one to build a ``reduction'' map from the set of values of $X_1, ..., X_n$ to the set of the partitions of $[n]$ as in the following example ($n=10$):
\begin{align}\label{gsh}
 \mathbb{N}^{10} &\rightarrow  \mathcal{P}_{[10]}  \\
X_1, ..., X_{10}  &\longmapsto  \Pi_{[10]}(X_1, X_2, ..., X_{10}) \\
(2, 4, 2, 4, 3, 3, 10, 13, 5, 4)&\longmapsto \{ \{1,3\}, \{2, 4, 10 \}, \{5, 6\}, \{7\}, \{8\}, \{9\} \} \end{align}

Similarly, and in agreement with Assumption~2, in our model we can consider the reduction of data which ignores information about the names of the DNA types: this is achieved, for instance, by retaining from the database only the equivalence classes of the indices of the individuals, according to the equivalence relation ``has the same DNA type''. Stated otherwise, the database is reduced to the partition $\pi_{[n]}^{\text{Db}}$, obtained using these equivalence classes. 
%Notice that the same partition is obtained via random variables $X_1, ..., X_n$, as defined in \eqref{eqx}.
%and further to the corresponding partition of $n$, denoted as $\pi_{n}^{\text{Db}}$. 
%Stated otherwise, we can reduce $B$ to $\pi_{[n]}^{\text{Db}}$, the partition of $[n]$ obtained from the database.
However, the database only supplies part of the data. There are also two new DNA profiles that are equal to one another (and different from the already observed ones in the rare type match case). 
Considering the suspect's profile we obtain the partition $\pi^{\text{Db}+}_{[n+1]}$, where the first $n$ integers are partitioned as in $\pi^{\text{Db}}_{[n]}$, and $n+1$ constitutes a class by itself. Considering the crime stain profile we obtain the partition $\pi^{\text{Db}++}_{[n+2]}$ where the first $n$ integers are partitioned as in  $\pi^{\text{Db}}_{[n]}$, and $n+1$ and $n+2$ belong to the same (new) class.
Random variables $\Pi^{\text{Db}}_{[n]}$, $\Pi^{\text{Db}+}_{[n+1]}$, and $\Pi^{\text{Db}++}_{[n+2]}$ are used to model $\pi_{[n]}^{\text{Db}}$, $\pi_{[n+1]}^{\text{Db}+}$, and $\pi_{[n+2]}^{\text{Db}++}$, respectively.

Since prosecution and defense agree on the distribution of $X_1, ..., X_{n+1}$, but not on the distribution of $X_{n+2}|X_1, ..., X_{n+1}$, they also agree on the distribution of $\Pi^{\text{Db}+}_{[n+1]}$ but disagree on the distribution of $\Pi^{\text{Db}++}_{[n+2]}$ (see \eqref{sdf}).

The crucial points of the model are the following:

\begin{enumerate}
\item the random partitions defined through random variables $X_1$, ..., $X_{n+2}$ and through database are the same.
\begin{align*}
\Pi_{[n]}^{\text{Db}}&= \Pi_{[n]}(X_1, ..., X_n),\\
\Pi_{[n+1]}^{\text{Db}+}&= \Pi_{[n+1]}(X_1, ..., X_{n+1}),\\
\Pi_{[n+2]}^{\text{Db}++}&= \Pi_{[n+2]}(X_1, ..., X_{n+2}).
\end{align*}
\item although $X_1$, ..., $X_{n+2}$ were not observable, the random partitions 
$\Pi_{[n]}^{\text{Db}}, \Pi_{[n+1]}^{\text{Db+}}$, and $\Pi_{[n+2]}^{\text{Db++}}$ are observable.
\end{enumerate}

To clarify, consider the following example of a database with $k=6$ different DNA types, from $n=10$ individuals: 
$$B=(b_1, b_2, b_1, b_2, b_3, b_3, b_4, b_5, b_6,b_2),$$ where $b_i$ is the name of the $i$th DNA type according to the order chosen for the database. 
This database can be reduced to the partition of $[10]$: $$\pi_{[10]}^{\text{Db}}=\{\{1,3\}, \{2, 4, 10\}, \{5,6\}, \{7\}, \{8\}, \{9\}\}.$$ 
Then, the part of reduced data whose distribution is agreed on by prosecution and defense is $$\pi_{[11]}^{\text{Db}+}=\{\{1,3\}, \{2, 4, 10\}, \{5,6\}, \{7\}, \{8\}, \{9\},  \{11\} \},$$ while the entire reduced data $D$ can be represented as  $$\pi_{[12]}^{\text{Db}++}=\{\{1,3\}, \{2, 4, 10\}, \{5,6\}, \{7\}, \{8\}, \{9\},  \{11, 12\} \}.$$

Now, assume that we know the rank in the population of each of the DNA types in the database: we know that $b_1$ is, for instance, the second most frequent type, $b_2$ is the fourth most frequent type, and so on.
Stated otherwise, we are now assuming that we observe the variables $X_1$, ..., $X_{n+2}$: for instance, $X_1= 2$,  $X_2=4$, $X_3=2$, $X_4=4$, $X_5= 3$, $X_6=3$, $X_7=10$, $X_8=13$, $X_9=5$, $X_{10}=4$, $X_{11}=9$, $X_{12}=9$ (as in \eqref{gsh}).
It is easy to check  that $\Pi_{[10]}(X_1, ..., X_{10})=\pi_{[10]}^{\text{Db}}$, $\Pi_{[11]}(X_1, ..., X_{11})=\pi_{[11]}^{\text{Db+}}$, and $\Pi_{[12]}(X_1, ..., X_{12})=\pi_{[12]}^{\text{Db++}}.$

Coming back to our model, data is  defined as $D=\pi^{\text{Db}++}_{[n+2]}$, obtained partitioning the database enlarged with the two new observations (or partitioning $X_1, ..., X_{n+2}$).
Node $D$ of Figure~\ref{berhnt} is defined accordingly.

Notice that, given $X_1,..., X_{n+2}$, $D$ is deterministic. An important result is that, according to Proposition 4 in \citet{pitman:1992b}, it is possible to derive directly the distribution of $D\mid \alpha, \theta, H$. In particular, it holds that if $$\mathbf{P}\mid \alpha, \theta \sim PD(\alpha, \theta),$$ and $$X_1, X_2, ... \mid \mathbf{P}=\mathbf{p} \sim_{\text{i.i.d}} \mathbf{p},$$ then, for all $n\in \mathbb{N}$, the random partition $\Pi_{[n]}=\Pi_{[n]}(X_1, ..., X_{n})$ has the following distribution:
\begin{equation}\label{ChEPPF}
\mathbb{P}_{n}^{\alpha,\theta}(\pi_{[n]}):=\Pr(\Pi_{[n]}=\pi_{[n]}| \alpha, \theta)=\frac{[\theta+ \alpha]_{k-1; \alpha}}{[\theta+ 1]_{n-1; 1}}\prod_{i=1}^k[1-\alpha]_{n_i-1;1}, 
\end{equation}
where $n_i$ is the size of the $i$th block of $\pi_{[n]}$ (the blocks are here ordered according to their least element), and $\forall x, b\in \mathbb{R}, a\in \mathbb{N}$, $[x]_{a,b}:=\begin{cases}
\prod_{i=0}^{a-1} (x+ib) &\text{if } a\in \mathbb{N}\backslash \{0\}\\
1 &\text{if } a=0
\end{cases}.$ This formula, also known as the \emph{Pitman sampling formula}, is further studied in \citet{pitman:1995} and shows that $\mathbb{P}_{n}^{\alpha,\theta}(\pi_{[n]})$ does not depend on $X_1, ..., X_{n}$, but only on the sizes and the number of classes in the partitions. It follows that we can get rid of the intermediate layer of nodes $X_1$, ..., $X_{n+2}$. Moreover, it holds have $\Pr(D|\alpha, \theta, h_p) = \mathbb{P}_{n+1}^{\alpha, \theta}(\pi_{[n+1]}^{\text{Db+}})$, while $\Pr(D|\alpha, \theta, h_d) = \mathbb{P}_{n+2}^{\alpha, \theta}(\pi_{[n+2]}^{\text{Db++}})$.

%

%
%Notice that to each partition $\pi_{n}= (n_1, ..., n_k)$ correspond $N^{\pi_n}$ different partitions of $[n]$, with $N^{\pi_n}:=\binom{n}{n_1, ..., n_k}\frac{1}{\prod_{i=1}^n a_i!}$, and $a_i=\#\{j| n_j=i\}$ \citep{pitman:1995}. Because of the symmetry, it holds that $\Pr_{\alpha,\theta}(\pi_n= (n_1, ..., n_k))= N^{\pi_n}p_{\alpha, \theta}(n_1, n_2, ..., n_k)$.
%

%For the moment, instead of specifying the distribution of $D^*$ we can just  re-write it in the following convenient form:
%\begin{align}
%&\Pr(D^*=\pi_{[n+2]}| \alpha, \theta)=\\
% &\Pr(\Pi_{[n+2]}=\pi_{[n+2]}| \Pi_{[n+1]}=\pi_{[n+1]}, \alpha, \theta)\Pr(\Pi_{[n+1]}=\pi_{[n+1]}| \alpha, \theta)
% \end{align}

\begin{figure}[htbp]
 
\centering
\begin{tikzpicture}  \node[draw, ellipse, minimum width=1.2cm]                (t) at (-2,0)  { $A, \Theta$};
\node [draw, ellipse, minimum width=1.2cm]              (h) at (-1,-1) { $H$};
 \node [draw, ellipse, minimum width=1.2cm]              (d) at (-2,-2) { $D$};

 \draw[black, big arrow] (t) -- (d);
 \draw[black, big arrow] (h) -- (d);     

\draw[black, big arrow] (t) -- (d);

\end{tikzpicture}
\caption{\small Simplified version of the Bayesian network in Figure~\ref{berhnt}}
\label{bernt2c}
 
\end{figure}
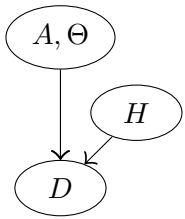

The model of Figure~\ref{berhnt} can thus be simplified to the one in Figure~\ref{bernt2c}. 

\subsection{Chinese Restaurant representation}\label{crp}
There is an alternative characterization of this model, called ``Chinese restaurant process'', due to \citet{aldous:1985} for the one-parameter case, and studied in detail for the two-parameter version in \citet{pitman:2006}. It is defined as follows:
consider a restaurant with infinitely many tables, each one infinitely large. 
Let $Y_1, Y_2, ...$ be integer-valued random variables that represent the seating plan: tables are ranked in order of occupancy, and $Y_i = j$ means that the $i$th customer seats at the $j$th table to be created. 
The process is described by the following transition matrix:
$$Y_1=1,$$
\begin{equation} \label{eqch}
\Pr(Y_{n+1} = i | Y_1, ..., Y_n)=\begin{cases} 
{\displaystyle \frac{\theta + k\alpha}{n+\theta} }& \text{if } i = k+1 \\
&\\
{\displaystyle\frac{n_i- \alpha}{n+\theta} }& \text{if } 1\leq i \leq k \\
\end{cases}
\end{equation}
where $k$ is the number of tables occupied by the first $n$ customers, and $n_i$ is the number of customers that occupy table $i$. The process depends on two parameters $\alpha$ and $\theta$ with the same conditions~\eqref{condition}. \textcolor{black}{From~\eqref{eqch} one can easily see the sufficientness property mentioned in Section~\ref{prior}.}

$Y_1, ..., Y_n$ are not i.i.d., nor exchangeable, but it holds that $\Pi_{[n]}(Y_1, ..., Y_n)$ is distributed as $\Pi_{[n]}(X_1, ..., X_n)$, with $X_1, ..., X_n$ defined as in \eqref{eqx}, and they are both distributed according to the Pitman sampling formula~\eqref{ChEPPF} \citep{pitman:2006}.

Stated otherwise, we can obtain $\pi_{[n]}^{\text{Db}}$ through seating plan of $n$ costumers or partitioning the $X_1, ..., X_n$ of the database and we obtain the same partition $\pi_{[n]}^{\text{Db}}$.
Similarly $\pi^{\text{Db}+}_{[n+1]}$ is obtained when a new customer has chosen an unoccupied table (remember we are in the rare type match case), and $\pi^{\text{Db}++}_{[n+2]}$ is obtained when the $(n+2)$nd customer goes to the table already chosen by the $(n+1)$st customer (suspect and crime stain have the same DNA type). In particular, thanks to \eqref{eqch}, we can write:
\begin{equation}\label{e1}
p(\pi_{[n+2]}^{\text{Db++}}\mid h_p, \pi_{[n+1]}^{\text{Db+}}, \alpha, \theta)=1,
\end{equation} 

\begin{equation}\label{e2}
p(\pi_{[n+2]}^{\text{Db++}}\mid h_d, \pi_{[n+1]}^{\text{Db+}}, \alpha, \theta)=\frac{1-\alpha}{n+1+\theta},
\end{equation}
since the $(n+2)$nd customer goes to the same table as the $(n+1)$st (who was sitting alone).

%Moreover, the map from $Y_1, ..., Y_n$ to $\Pi_{[n]}:=\Pi_{[n]}(Y_1, ..., Y_n)$ is a bijection: from the partition it is possible to go back to the observation $Y_1,...,Y_n$.

%In the forthcoming section we will make use of the following notation:
%$K_n:= |\Pi_{[n]}|$, the number of tables occupied at time $n$.

\subsection{A useful Lemma}\label{gm}

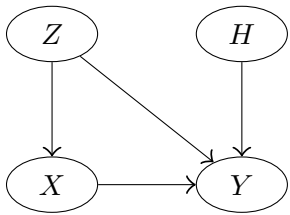
\begin{figure}[htbp]
 
\centering
  \begin{tikzpicture}
\node[draw, ellipse, minimum width=1.2cm]                (t) at (-1,0)  { $Z$};
\node [draw,  ellipse, minimum width=1.2cm]              (h) at (1.5,0) {$H$};
 \node [draw, ellipse, minimum width=1.2cm]              (d) at (-1,-2) { $X$};
\node [draw, ellipse, minimum width=1.2cm]              (dr) at (1.5,-2) { $Y$};
  \draw[black, big arrow]  (t) -- (d);
  \draw[black, big arrow] (h) -- (dr);     
  \draw[black, big arrow]  (t) -- (dr);
   \draw[black, big arrow] (d) -- (dr);
\end{tikzpicture}

     \caption{\small Conditional dependencies of the random variables of Lemma~\ref{lemma1}}\label{giz1}
   
     \end{figure}

The following lemma can be applied to four general random variables $Z$, $X$, $Y$, and $H$ whose conditional dependencies are described by the Bayesian network of Figure~\ref{giz1}. The importance of this result is due to the possibility of using it \textcolor{black}{ for assessing the likelihood ratio} in a very common forensic situation: the prosecution and the defense disagree on the distribution of the entirety of data ($Y$) but agree on the distribution of a part it ($X$), and these distributions depend on parameters ($Z$). 

   \begin{lemma}\label{lemma1}
Given four random variables $Z$, $H$, $X$ and $Y$, whose conditional dependencies are represented by the Bayesian network of Figure~\ref{giz1}, the likelihood function for $h$, given $X=x$ and $Y=y$ satisfies 
$$\mathrm{lik}(h\mid x, y)~ \propto \mathbb{E}(p(y \mid x, Z, h) \mid X = x).$$   \end{lemma}

The Bayesian representation of the model, in Figure~\ref{giz1}, allow to factor the joint probability density of $Z$,  $H$, $X$ and $Y$ as
$$p(z, h, x, y) ~=~ p(z) \, p(x \mid z)\, p(h)\, p(y \mid x, z, h).$$

\noindent By Bayes formula, $p(z) \, p(x \mid z)= p(x)\,p(z\mid x)$. This rewriting corresponds to reversing the direction of the arrow between $Z$ and $X$:
     
\bigskip     

\begin{center}
\begin{tikzpicture}
\node [draw, ellipse, minimum width=1.2cm]              (z) at (-1,0)  { $Z$};
\node [draw, ellipse, minimum width=1.2cm]              (h) at (1.5,0) {$H$};
\node [draw, ellipse, minimum width=1.2cm]              (x) at (-1,-2) {$X$};
\node [draw, ellipse, minimum width=1.2cm]              (y) at (1.5,-2) { $Y$};

\draw [black, big arrow] (x) -- (z);
\draw [black, big arrow] (z) -- (y);     
\draw [black, big arrow] (h) -- (y);
\draw [black, big arrow] (x) -- (y);
\end{tikzpicture}
\end{center}
\noindent The random variable $X$ is now a root node. This means that when we probabilistically condition on $X=x$, the graphical model changes in a simple way: we can delete the node $X$, but just insert the value $x$ as a parameter in the conditional probability tables of the variables $Z$ and $Y$ which formerly had an arrow from node $X$. The next graph represents this model:
\bigskip

\begin{center}
\begin{tikzpicture}
\node [draw, ellipse, minimum width=1.2cm]              (a) at (-1,0)  { $Z$};
\node [draw, ellipse, minimum width=1.2cm]              (h) at (1.5,0) {$H$};
\node                                                                    (x1) at (-1,-1.5) { $x$};
\node                                                                    (x2) at (0,-2) { $x$};
\node [draw, ellipse, minimum width=1.2cm]              (y) at (1.5,-2) { $Y$};

\draw [black, dash arrow, dashed] (x1) -- (z);
\draw [black, big arrow] (z) -- (y);     
\draw [black, big arrow] (h) -- (y);
\draw [black, dash arrow, dashed] (x2) -- (y);
\end{tikzpicture}
\end{center}
\noindent This tells us, that conditional on $X=x$, the joint density of $Z$, $Y$ and $H$ is equal to 
$$p(z,h,y \mid x)=p(z\mid x) p(h) p(y \mid x, z, h).$$
The joint density of $H$ and $Y$ given $X$ is obtained by integrating out the variable $Z$. It can be expressed as a conditional expectation value since $p(z\mid x)$ is the density of $Z$ given $X=x$. We find:
$$p(h,y \mid x) = p(h) \mathbb{E}(p(y \mid x, Z, h) \mid X = x).$$

Recall that this is the joint density of two of our variables, $H$ and $Y$, after conditioning on the value $X=x$. Let us now also condition on $Y=y$. It follows that the density of $H$ given $X=x$ and $Y=y$ is proportional (as function of $H$, for fixed $x$ and $y$) to the same expression, $p(h) \mathbb{E}(p(y \mid x, Z, h) \mid X= x)$. 

This is a product of the prior for $h$ with some function of $x$ and $y$. Since posterior odds equals prior odds times likelihood ratio, it follows that the likelihood function for $h$, given $X=x$ and $Y=y$ satisfies
$$\textrm{lik}(h\mid x, y)~ \propto \mathbb{E}(p(y \mid x, Z, h) \mid X = x).$$

   \begin{corollary}\label{cor1}
Given four random variables $Z$, $H$, $X$ and $Y$, whose conditional dependencies are represented by the network of Figure~\ref{giz1}, the likelihood ratio for $H=h_1$ against $H=h_2$ given $X=x$ and $Y=y$ satisfies 
         \begin{equation}\label{LR_fo}
         \mathrm{LR}=\frac{\mathbb{E}(p(y|x, Z, h_1)|X=x)}{\mathbb{E}(p(y|x, Z, h_2)|X=x)}.
         \end{equation}
   \end{corollary}

\section{The likelihood ratio}\label{lr}

\textcolor{black}{From now on we will omit the superscripts Db, Db+, and Db++ for ease of notation.}

Using the hypotheses and the reduction of data $D$ defined in Section~\ref{m}, the likelihood ratio will be defined as $$\text{LR}= \frac{p(\pi_{[n+2]}|h_p)}{p(\pi_{[n+2]}|h_d)}= \frac{p(\pi_{[n+1]},\pi_{[n+2]}|h_p)}{p(\pi_{[n+1]},\pi_{[n+2]}|h_d)}.$$
The last equality holds due to the fact that $\Pi_{[n+1]}$ is a deterministic function of $\Pi_{[n+2]}$. 

Corollary~\ref{cor1} can be applied to our model since defense and prosecution agree on the distribution of $\pi_{[n+1]}$, but not on the distribution of $\pi_{[n+2]}$, and data depends on parameters $\alpha$ and $\theta$.
Thus, if $(A,\Theta)$ play the role of $Z$, $X= \Pi_{[n+1]}$, and $Y= \Pi_{[n+2]}$, by using \eqref{e1} and \eqref{e2}, we obtain:
   \[
\begin{aligned}\label{ffd}
\text{LR}~&=\frac{\mathbb{E}(p(\pi_{[n+2]}\mid \pi_{[n+1]}, A, \Theta, h_p ) \mid \Pi_{[n+1]}= \pi_{[n+1]}) }{ \mathbb{E}(p(\pi_{[n+2]}\mid \pi_{[n+1]}, A, \Theta, h_d ) \mid \Pi_{[n+1]}= \pi_{[n+1]}) }\\[11pt]
&=\frac{1}{\mathbb{E}\Big(\frac{1-A}{n+1+\Theta}\mid \Pi_{[n+1]}= \pi_{[n+1]}\Big)}.
\end{aligned}
\]
The expected value is taken with respect to the posterior distribution of $A, \Theta\mid \Pi_{[n+1]}= \pi_{[n+1]}$.
\textcolor{black}{The solution we propose in this paper is to deal with the uncertainty about $\alpha$ and $\theta$ by using MLE estimators and plug those estimators into the formula. 
Notice that this is equivalent to a hybrid approach, in which the parameters are estimated in a frequentist way and their values are plugged into the Bayesian LR. In the future, we plan to use MCMC methods to calculate as exactly as possible the exact posterior distribution, given assumed priors on the hyperparameters.} 

By defining the random variable ${\displaystyle\Phi=n\frac{1-A}{n+1+\Theta}}$ we can write the LR as
\begin{equation*}\label{ffd2}
\text{LR}=\frac{n}{\mathbb{E}(\Phi \mid \Pi_{[n+1]}=\pi_{[n+1]})}.
\end{equation*}
\section{Analysis on YHRD database}\label{ard}
In this section, we present the study we made on a database of 18,925 Y-STR 23-loci profiles from 129 different locations in 51 countries in Europe \citep{purps:2014}\footnote{The database has previously been cleaned by Mikkel Meyer Andersen (\url{http://people.math.aau.dk/~mikl/?p=y23}).}.
Our analyses are performed by considering only 7 Y-STR loci (DYS19, DYS389 I, DYS389 II, DYS3904, DYS3915, DY3926, DY3937) but similar results have been observed with the use of 10 loci.

\subsection{Model fitting}\label{fit}
First, we calculated the maximum likelihood estimators $\alpha_{\text{MLE}}$ and $\theta_{\text{MLE}}$ using the entire database and the likelihood defined by \eqref{ChEPPF}. Their values are $\alpha_{\text{MLE}}=0.51$ and $\theta_{\text{MLE}}=216.$

\begin{figure}[htbp]%
 
\centering
\includegraphics[scale=0.4]{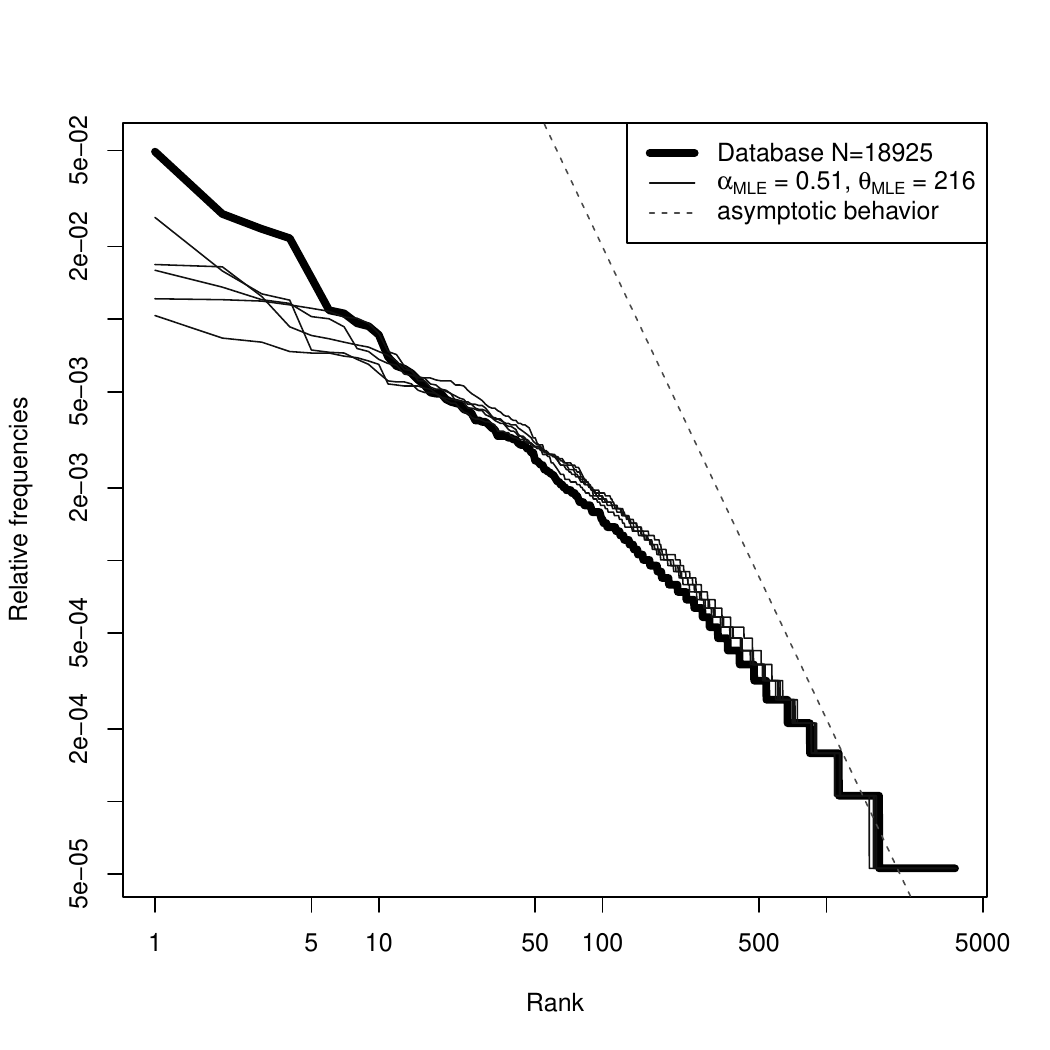}
\caption{\small Log scale ranked frequencies from the database (thick line) are compared to the relative frequencies of samples of size $n=18'925$ obtained from several realizations of PD($\alpha_{MLE}, \theta_{MLE}$) (thin lines). \textcolor{black}{Asymptotic power-law behavior is also displayed (dotted line)}.}\label{3figs}
 
\end{figure}

In Figure~\ref{3figs}, the ranked frequencies of the 18'925 Y-STR profiles of the YHRD database are compared to the relative frequencies of samples of size $n$ obtained from several realizations of PD($\alpha_{MLE}, \theta_{MLE}$).
To do so we run several times the Chinese Restaurant seating plan (up to $n=18,925$ customers): each run is \textcolor{black}{used to approximate} a new realization $\mathbf{p}$ from the PD($\alpha_{MLE}, \theta_{MLE}$). As explained in Section \ref{crp}, the partition of the customers into tables is the same as the partition obtained from an i.i.d.\ sample of size $n$ from $\mathbf{p}$. We can see that for the most common haplotypes (left part of the plot) there is some discrepancy. However, we are interested in rare haplotypes, which typically have a frequency belonging to the right part of the plot. In that region, the two-parameter Poisson Dirichlet follows the distribution of the data quite well. 
\textcolor{black}{
The dotted line shows in Figure~\ref{3figs} the asymptotic behavior on the two-parameter Poisson Dirichlet distribution. Indeed, if $\mathbf{P} \sim \text{PD}(\alpha, \theta)$, then
\begin{equation*} 
\frac{P_i}{Z i^{-1/\alpha}}\rightarrow 1, \quad \text{a.s., when  } i\rightarrow +\infty
\end{equation*} 
for a random variable $Z$ such that $Z^{-\alpha}= \Gamma(1-\alpha)/S_{\alpha}$. This power-law behavior describes an incredible variety of phenomena \citep{newman:2005}. }

\begin{figure}[htbp]%
 
\centering
\includegraphics[scale=0.4]{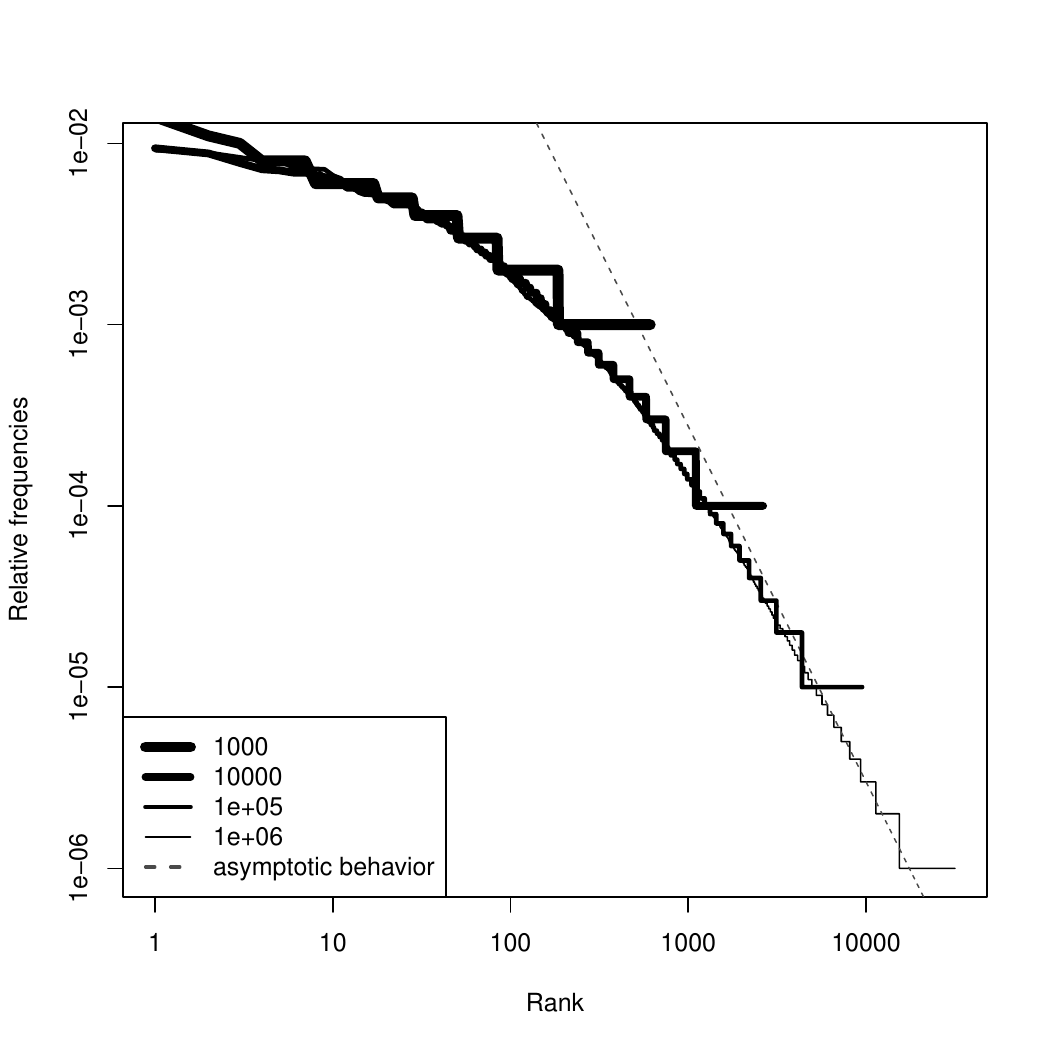}
\caption{\small Log scale ranked frequencies from the two-parameter Poisson Dirichlet distribution with $\alpha=0.51, \theta=216$ approximated through a Chinese restaurant seating plan, each with its ownnumber of costumers, corresponding to the different thickness of the lines.}\label{3figsd}
 
\end{figure}

\textcolor{black}{The thick line in Figure~\ref{3figs} also seems to have a power-law behavior, and to be honest, we were hoping to get the same asymptotic slope of the prior.This is not what we observe, but in Figure~\ref{3figsd} it can be seen that for such a big value of $\theta$ we would need a bigger database (at least $n=10^6$) to see the correct slope.
 }
 
  \subsection{Log-likelihood}
\begin{figure}[htbp]
 
\centering
%\subfigure[Relative loglikelihood for parameters $\alpha$ and $\theta$, compared to a Gaussian distribution, $95\%$ and $99\%$ confidence intervals (black and red) ]{
%\includegraphics[scale=0.5,width=0.45\textwidth ]{Figure_due_paper.pdf}
%}

\includegraphics[scale=0.5,width=0.45\textwidth]{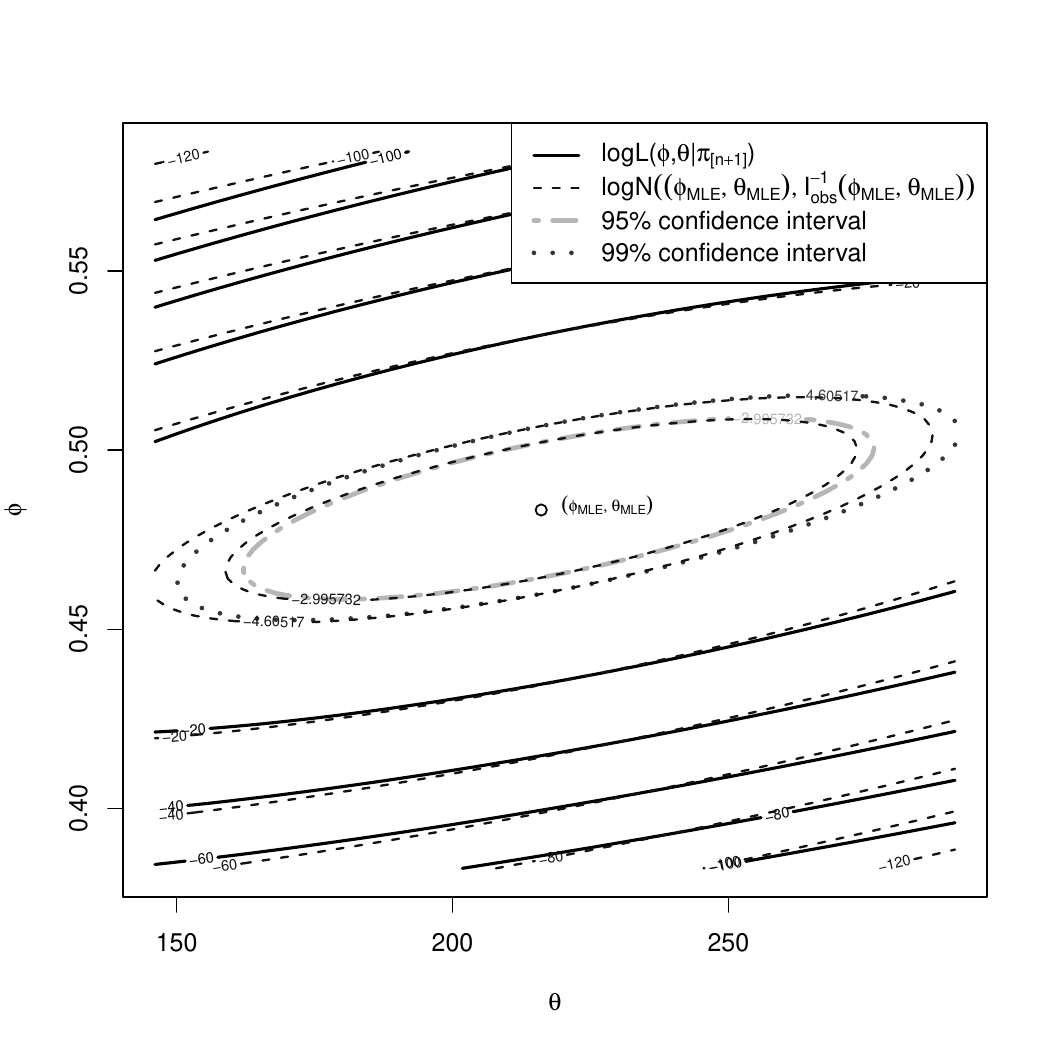}

\caption{\small Relative log-likelihood for $\phi=n\frac{1-\alpha}{ n+1+\theta}$ and $\theta$ compared to a Gaussian distribution displayed with $95\%$ and $99\%$ confidence intervals}\label{2figs}
 
\end{figure}

It is also interesting to investigate the shape of the log-likelihood function for $\alpha$ and $\theta$ given $\pi_{[n+1]}$. It is defined as 

$$l_{n+1}(\alpha, \theta):=\log p(\pi_{[n+1]}| \alpha, \theta).$$
%=\frac{[\theta+ \alpha]_{k ; \alpha}}{[\theta+ 1]_{n; 1}}\prod_{i=1}^{k+1}[1-\alpha]_{n_i-1;1}.$$
%
%In Figure~\ref{2figs} (a), the loglikelihood function is compared to the Gaussian distribution centered in the maximum likelihood estimates for $\alpha$ and $\theta$, with the inverse of the observed Fisher information as covariance matrix. 
In Figure~\ref{2figs} the log-likelihood reparametrized using $\phi={\displaystyle n\frac{1-\alpha}{n+1+\theta}}$ instead of $\alpha$ is displayed. A Gaussian distribution centered in the MLE parameters and with covariance matrix the inverse of the Fisher Information, is also displayed (in dashed lines). This is not done to show an asymptotic property, but to show the symmetry of the log-likelihood, which \textcolor{black}{validates} approximation of $\mathbb{E}(\Phi \mid \Pi_{[n+1]}=\pi_{[n+1]})$ with the marginal mode $\Phi_{MLE}$, at least when we choose an hyperprior $p(\phi, \theta)$ that is flat around $(\phi_{MLE}, \theta_{MLE})$: indeed, it holds that $p(\phi, \theta \mid \pi_{[n+1]})\propto l_{n+1}(\phi, \theta) \times p(\phi, \theta)$.

Hence, one could safely make this approximation if one believed that this symmetry would also be true in the real data situation at hand:
\begin{equation} \label{mlelr}
\text{LR}\approx \frac{n+1+\theta_{MLE}}{1-\alpha_{MLE}}.
\end{equation}

Notice that this is equivalent to a hybrid approach, commonly called ``Empirical Bayes'', in which the parameters are estimated through the MLE (frequentist) and their values are plugged into the Bayesian LR.
\textcolor{black}{We would like to reiterate that we are not using maximum likelihood estimates of the parameters because we consider the likelihood ratio from a frequentist point of view. Our aim is to calculate a Bayesian likelihood ratio, and we have observed empirically that using the maximum likelihood estimates of the parameters we can approximate this value. }

Hence, in case of a rare type match problem, and using the YHRD database as the reference database, we have $\log_{10}\LR=4.59$, that corresponds to say that it is approximately 40,000 times more likely to observe the reduced data under the prosecution hypothesis than under the defence hypothesis.

 \subsection{True LR}\label{tlr}

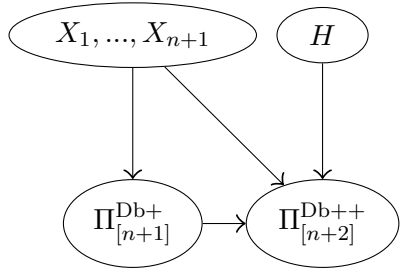
\begin{figure}[htbp]
 
\centering
  \begin{tikzpicture}
\node[draw, ellipse, minimum width=1.2cm]                (t) at (-1,0)  { $X_1, ..., X_{n+1} $};
\node [draw, ellipse, minimum width=1.2cm]              (h) at (1.5,0) {$H$};
 \node [draw, ellipse, minimum width=1.2cm]              (d) at (-1,-2.5) { $\Pi_{[n+1]}^{\text{Db}+}$};
\node [draw, ellipse, minimum width=1.2cm]              (dr) at (1.5,-2.5) { $\Pi_{[n+2]}^{\text{Db}++}$};
  \draw[black, big arrow]  (t) -- (d);
  \draw[black, big arrow] (h) -- (dr);     
  \draw[black, big arrow]  (t) -- (dr);
   \draw[black, big arrow] (d) -- (dr);
\end{tikzpicture}

     \caption{\small Bayesian network for the case in which $\mathbf{p}$ is known.}\label{fig_bnet_due}
     
     \end{figure}

 It is also interesting to study the  likelihood ratio values obtained with out method according to formula \eqref{ffd2}, and to compare it with the `true' ones, meaning the LR values obtained when the vector $\mathbf{p}$ is known, over simulated rare type match cases. This corresponds to the desirable (even though completely imaginary) situation of knowing the ranked list of the frequencies of all the DNA types in the population of interest. The model can be represented by the Bayesian network of Figure~\ref{fig_bnet_due}.
 
 The likelihood ratio in this case can be obtained using again Corollary~\ref{cor1}, where now $X_1$, ..., $X_{n+1}$ play the role of $Z$.
 Indeed, \textcolor{black}{now that $\mathbf{p}$ is known, the unobservable part of the model are the ranks of the types in the database.}
   \begin{align*}
 \text{LR}_{\mid \mathbf{p}}&=\frac{p(\pi_{[n+2]}, \pi_{[n+1]} \mid h_p, \mathbf{p})}{p(\pi_{[n+2]}, \pi_{[n+1]}\mid h_d, \mathbf{p})}\\
 &= \frac{\mathbb{E}(p(\pi_{[n+2]} \mid  \pi_{[n+1]}, X_1, ..., X_{n+1}, h_p, \mathbf{p})\mid \Pi_{[n+1]} =  \pi_{[n+1]}, \mathbf{p} )}{\mathbb{E}(p(\pi_{[n+2]}\mid  \pi_{[n+1]}, X_1, ..., X_{n+1}, h_d, \mathbf{p})\mid \Pi_{[n+1]} =  \pi_{[n+1]}, \mathbf{p})}\\
&= \frac{1}{\mathbb{E}(p_{X_{n+1}}|\Pi_{[n+1]}=\pi_{[n+1]}, \mathbf{p})}.
 \end{align*}

Notice that, in the rare type case, $X_{n+1}$ is observed only once among the $X_1$, ..., $X_{n+1}$. Hence, we call it a singleton, and its distribution given $\mathbf{p},\pi_{[n+1]}$ is the same as the distribution of each other singleton. Let $s_1$ denote the number of singletons, and $\mathcal{S}$ the set of indices of singletons observations in the \textcolor{black}{augmented} database. It holds 
$$
s_1 \mathbb{E}(p_{X_{n+1}}|\pi_{[n+1]}, \mathbf{p}) =  \mathbb{E}(\sum_{i \in \mathcal{S}}p_{X_{i}}|\pi_{[n+1]}, \mathbf{p}).
$$

Notice also that the knowledge of $\mathbf{p}$ and $\pi_{[n+1]}$, is not enough to observe $X_1, ..., , X_{n+1}$. 

Let us denote as $X_1^*$, .., $X^*_K$ the $K$ different values taken by $X_1$, ..., $X_{n+1}$,  ordered decreasingly according to the frequency of their values. Stated otherwise, if $n_i$ is the frequency of $x_i^*$ among $x_1, ...,  x_{n+1},$ then $n_1\geq n_2\geq ... \geq n_K$. \footnote{Moreover, in case $X_i^*$ and $X_j^*$ have the same frequency ($n_i=n_j$), then they are ordered increasingly according to their values.} For instance, if $X_1= 2$,  $X_2=4$, $X_3=2$, $X_4=4$, $X_5= 3$, $X_6=3$, $X_7=10$, $X_8=13$, $X_9=5$, $X_{10}=4$, $X_{11}=9$, then 
$X_1^*= 4, X_2^*=2, X_3^*=3, X_4^*=5, X_5^*=9, X_6^*=10, X_7^*=13$.

By definition, it holds that
$$\mathbb{E}(\sum_{i \in \mathcal{S}}p_{X_{i}}|\pi_{[n+1]}, \mathbf{p}) = \mathbb{E}(\sum_{j: \, n_j=1}p_{X_{j}^*}|\pi_{[n+1]}, \mathbf{p}).
$$

Notice that $(n_1, n_2, ..., n_K)$ is a partition of $n+1$, which will be denoted as $\pi_{n+1}$. In the example, $\pi_{n+1}=(3, 2, 2, 1, 1, 1, 1)$. 
Since the distribution of ${\displaystyle \sum_{j: \, n_j=1}p_{\textcolor{black}{X_{j}^*}}}$ only depends on $\pi_{n+1}^{\text{Db+}}$, the latter can replace $\pi_{[n+1]}^{\text{Db+}}$. Thus, it holds that
\begin{equation}\label{lrp}
\textrm{LR}_{\mid\mathbf{p}}=\frac{s_1}{\mathbb{E}({\displaystyle \sum_{j: \, n_j=1}}p_{X_{j}^*}|\pi_{n+1}^{\text{Db+}}, \mathbf{p})}.
\end{equation}

%
%\sout{For the same reason explained above, knowledge of $\mathbf{p}$ and $\pi_{n+1}^{\text{Db+}}$ is not enough to observe $X_1^*, ..., X_K^*$.}
A more compact representation for $\pi_{n+1}^{\text{Db+}}$ can be obtained by using two vectors $\mathbf{a}$ and $\mathbf{r}$ where $a_j$ are the
distinct numbers occurring in the partition, increasingly ordered, and each $r_j$ is the number of
repetitions of $a_j$. $J$ is the length of these two vectors, and it holds that $n+1=\sum_{j=1}^J a_j r_j.$
In the example above we have that $\pi_{n+1}$ can be represented by $(\mathbf{a}, \mathbf{r})$ with $\mathbf{a}=(1,2,3)$ and $\mathbf{r}=(4, 2,1), J=3$.

There is an unknown map, $\chi$, treated here as latent variable, which assigns the ranks of the DNA types, ordered according to their frequency in Nature, to one of the number $ \{1, 2, ..., J\}$ corresponding to the position in $\mathbf{a}$ of its frequency in the sample, or to $0$ if the type if not observed. Stated otherwise, $$ \chi: \{1, 2, ...\} \longrightarrow \{0,1, 2, ..., J\}$$
$$\chi(i)=\begin{cases}
0 & \textrm{if the $i$th most common species in Nature is not observed in the sample},\\
j & \textrm{if the $i$th most common species in Nature is one of the $r_j$ observed $a_j$ times in the sample.}
\end{cases}$$
 
 Given $\pi_{n+1}=(\mathbf{a}, \mathbf{r})$, $\chi$ must satisfy the following set of $J$ conditions:
 \begin{equation}\label{condfi}
 \sum_{i=1}^{\infty} \mathbf{1}_{\chi(i)=j}=r_j, \qquad \forall j\in \{1,..., J\}.
 \end{equation}
 
\textcolor{black}{In addition, it should not be allowed that a profiles observed $k_N$ times in the population is observed $k_n >k_N$ times in the sample. Hence we have to add a further condition:}
  \begin{equation}\label{condfi2}
N p_i > a_{\chi(i)}, \qquad \forall i
 \end{equation}
\textcolor{black}{where $N$ is the size of the entire population.}

The map $\chi$ can be represented by a vector $\boldsymbol{\chi}=(\chi_1, \chi_2, ...)$ such that $\chi_i=\chi(i)$. 
In the example  we have that $$\boldsymbol{\chi}=(0, 2, 2, 3, 1, 0, 0, 0, 1, 1, 0, 0, 1, 0, 0,...).$$

Notice that, given $\pi_{n+1}=(\mathbf{a},\mathbf{r})$, the knowledge of $\boldsymbol{\chi}$ implies the knowledge of $X_1^*$, ..., $X_K^*$: indeed it is enough to consider the position of the ranked positive values of $\boldsymbol{\chi}$, and to solve ties by considering the positions themselves (if $\chi_i=\chi_j$, than the order is given by $i$ and $j$). 
For instance, in the example if we sort the positive values of $\boldsymbol{\chi}$ and we collect their positions we get
$(4, 2, 3, 5, 9,10, 13)$: the reader can notice that we got back to $X_1^*, ..., X_7^*$.

This means that to obtain the distribution of $X_1^*,..., X_{K}^*| \pi_{n+1}, \mathbf{p}$, which appears in~\eqref{lrp}, it is enough to obtain the distribution of $\boldsymbol{\chi}|\mathbf{a},\mathbf{r}, \mathbf{p}$, and since we are only interested in the mean of the sum of singletons in samples of size $n+1$ from the distribution of $X_1^*,..., X_{K}^*|\mathbf{a},\mathbf{r}, \mathbf{p}$, we can just simulate samples from the distribution of $\boldsymbol{\chi}|\mathbf{a},\mathbf{r}, \mathbf{p}$ and sum the $p_i$ such that $\chi_i=1.$

\textcolor{black}{It holds that}
\textcolor{black}{\begin{equation}\label{23}
p(\mathbf{a},\mathbf{r} \mid \boldsymbol{\chi}, \mathbf{p} )\propto  \prod_{1\leq i\leq m} p_i^{a_{\chi_i}}, \end{equation}}
\textcolor{black}{where the proportionality factor is $\frac{(n+1)!}{\prod_{1\leq j\leq J}(a_j!)^{r_j}}$.}

\paragraph{Details of the Metropolis Hashting algorithm}

\textcolor{black}{Notice that for the model we assumed $\mathbf{p}$ to be infinitely long, but for simulations we will use a finite $\mathbf{\bar{p}}$, of length $m$. This is equivalent to assume that only $m$ elements in the infinite $\mathbf{p}$ are positive, and the remaining infinite tail is made of zeros. }

\textcolor{black}{To simulate samples from the distribution of $\boldsymbol{\chi}|\mathbf{a},\mathbf{r}, \mathbf{p}$ we use a Metropolis-Hastings algorithm on the space of the vectors $\boldsymbol{\chi}$ satisfying the $J+m$ conditions~\eqref{condfi} and ~\eqref{condfi2}.
Then the state space of the Metropolis-Hastings Markov chain is made of all vectors of length $m$ whose elements belong to $\{0, 1, ..., J\}$, and satisfy the conditions~\eqref{condfi} and~\eqref{condfi2}. 
If we start with an initial point $\boldsymbol{\chi}_{0}$ which satisfies~\eqref{condfi} and, at each move $t$ of the Metropolis-Hastings we swap two different values $\chi_i$ and $\chi_j$ inside the vector, condition \eqref{condfi} remains satisfied \textcolor{black}{while conditions~\eqref{condfi2} must be checked at every iterations. }
The Metropolis factor is the ratio of the two likelihoods $p(\mathbf{a},\mathbf{r}\mid \boldsymbol{\chi}_t , \mathbf{p} ) $ and $p(\mathbf{a},\mathbf{r} \mid \boldsymbol{\chi}_{ t+1},  \mathbf{p})$ where $\boldsymbol{\chi}_{ t}$ and $\boldsymbol{\chi}_{ t+1}$ differs only because $\chi_i$ and $\chi_j$ are exchanged. Hence, using \eqref{23}, the Metropolis factor for every move is
$$
R= \frac{p_i^{a_{\chi_i}} p_j^{a_{\chi_j}}}{p_j^{a_{\chi_i}}p_i^{a_{\chi_j}}}
$$. 
Every exchange move is then accepted with probability $R$. The algorithm is iterated $N=10^5$ times, with thinning steps of $10^3$ and a burnin period of 20000 iterations. 
Since it holds that
$$\mathbb{E}({\displaystyle \sum_{j: \, n_j=1}}p_{X_{j}^*}|\pi_{n+1}^{\text{Db+}}, \mathbf{p})= \mathbb{E}({\displaystyle \sum_{i :\chi_i=1}}p_{i}| \mathbf{a}, \mathbf{r}, \mathbf{p}),$$ 
for every accepted $\boldsymbol{\chi}$ we calculate the sum of all $p_i$s such that $\chi_i=1$ and we use the average to approximate the denominator of \eqref{lrp}. }

The algorithm is based on a similar one proposed in~\citet{anevski:2013}.

This method allows us to approximate the `true' LR when the vector $\mathbf{p}$ is known. This is almost never the case, but we can put ourselves in a fictitious world where we know $\mathbf{p}$ (such as the frequencies in the YHRD database, or as in the following section the frequencies from a smaller population)  and compare the true values for the $\LR_{\mid\mathbf{p}}$ with the one obtained by applying our Bayesian nonparametric model when $\mathbf{p}$ is unknown. 

\subsection{Frequentist-Bayesian analysis of the error}

A real Bayesian statistician chooses the prior and hyperprior according to his beliefs. Depending on the choice of the hyperprior over $\alpha$ and $\theta$ she may or may not believe in the approximation~\eqref{mlelr}, but she does not really talk of `error'.
However, hardliner Bayesian statisticians are a rare species, and most of the time the Bayesian procedure consists in choosing priors (and hyperpriors) which are a compromise between personal beliefs and mathematical convenience.
It is thus interesting to investigate the performance of such priors. This can be done by comparing the Bayesian likelihood ratio with the likelihood ratio that one would obtain if the vector $\mathbf{p}$ was known, and for the same reduction of data. This is what we call `error': in other words, at the moment we are considering the Bayesian nonparametric method proposed in this paper as a way to estimate (notice the frequentist terminology) the true $\LR_{\mid\mathbf{p}}$. If we denote by $p_x$ the population proportion of the matching profile, another interesting comparison is the one between the Bayesian likelihood ratio and the frequentist likelihood ratio $1/p_x$ (here denoted as $\LR_f$) that one would obtain knowing $\mathbf{p}$, but not reducing the data to partition. This is a sort of benchmark comparison and tells us how much we lose by using the Bayesian nonparametric methodology, and by reducing data. 

In total there are three quantities of interest ($\log_{10}\LR$, $\log_{10}\LR_{\mid\mathbf{p}}$, and $\log_{10}\LR_f$), and two differences of interest, which will be denoted as
\begin{itemize}
\item $\textrm{Diff}_1=\log_{10}\LR - \log_{10}\LR_{\mid\mathbf{p}}$ (loss due to choice of the Poisson Dirichlet model and approximation \eqref{mlelr}),
\item $\textrm{Diff}_2=\log_{10}\LR - \log_{10}\LR_{f}$ (overall loss).
\end{itemize}

In order to analyse these five quantities, we can study their distribution over different rare type match cases.
However, there is an obstacle. The Metropolis-Hastings algorithm described in Section~\ref{tlr} is too slow to be used with the entire European database of \citet{purps:2014} of size $n=18'925$.
%Further investigations can be carried out to find out whether the approximation \eqref{mlelr} is meaningful for smaller databases (for instance subsamples from the YHRD data). 
%Once this is checked the entire European database can be used as a ground-truth population from which many samples of smaller size are drawn and used as reference databases available for the case ($B$).  
%This would allow to simulate a distribution for the three differences described above.

In order to make the computational effort feasible, we consider the haplotype frequencies for the sole Dutch population (of size $n=2037$), and we pretend that they are the frequencies from the entire population of possible perpetrators.  This population is summarised by the following $\mathbf{a}$, and $\mathbf{r}$:

\small{
$$\mathbf{a}=(1 , 2 , 3 , 4 , 5 , 6 , 7 , 8 , 9 , 10 , 11 , 12 , 14 , 15 , 16,17 , 19 , 20 , 23 , 24 , 29 , 35 , 41 , 46 , 94 , 152 , 168 , 174)$$
$$\mathbf{r}=(356 ,  80 ,  31 ,  20 ,  13 ,  11 ,   5 ,   6 ,   3 ,   5 ,   4 ,   3 ,   2 ,   3 ,1, 1 ,   1 ,   1 ,   1 ,   1 ,   1 ,   1 ,   1 ,   1 ,   2 ,   1 ,   1 ,   1 ),$$
}
\normalsize
 and the maximum likelihood estimators for $\alpha$ and $\theta$ are $\alpha_{\text{MLE}}=0.62$, $\theta_{\text{MLE}}=22$.

%\begin{table}[htbp]
%\centering
%\begin{tabular}{|l|rrrrrrrrrrrrrrrr|}
%  \hline
% Freqs& 1 & 2 & 3 & 4 & 5 & 6 & 7 & 8 & 9 & 10 & 11 & 12 & 14 & 15 & 16&\\
%  
%  \hline
% Reps&356 &  80 &  31 &  20 &  13 &  11 &   5 &   6 &   3 &   5 &   4 &   3 &   2 &   3 &1&\\
%  \hline
% & \\
%     \hline
%   Freqs&17 & 19 & 20 & 23 & 24 & 29 & 35 & 41 & 46 & 94 & 152 & 168 & 174&& &\\
%    \hline
%    Reps  && &\\ 
%   \hline
%\end{tabular}
%    \caption{Summary of the frequencies of the Dutch population. For instance, Freqs=1 and Reps=356 means that there are 356 singletons, and 80 duplets, etc. }\label{dutch}
%\end{table}

\begin{figure}[htbp]%
 
\centering
\includegraphics[scale=0.4]{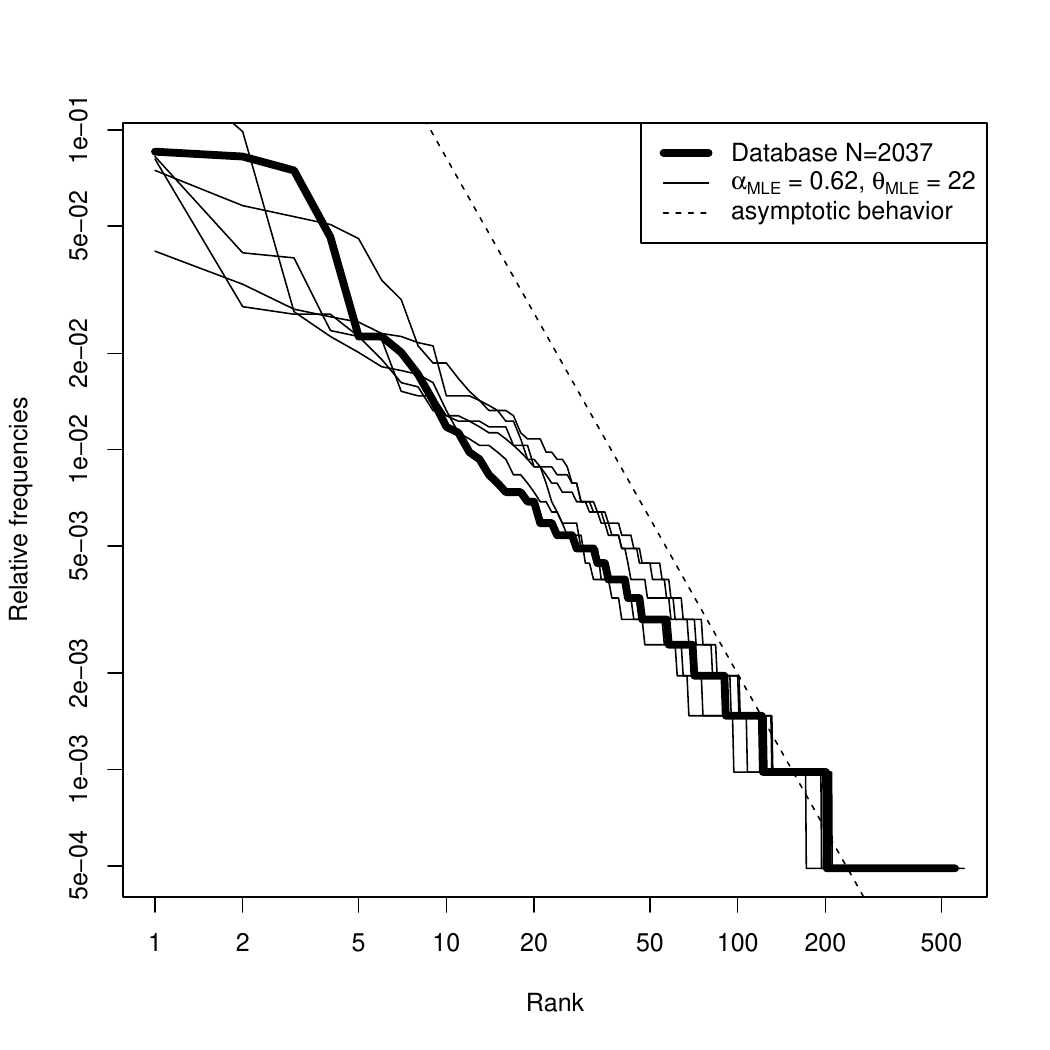}
\caption{\small Log scale ranked frequencies from the Dutch database (thick line)  compared to the relative frequencies of samples of size $n=2037$ obtained from several realizations of PD($\alpha_{MLE}, \theta_{MLE}$) (thin lines). \textcolor{black}{Asymptotic power-law behavior is also displayed (dotted line)}.}\label{3fiegs}
 
\end{figure}

In this way we can use the Metropolis-Hashting algorithm to simulate $\LR_{\mid \mathbf{p}}$. The model fitting is still good enough, as shown in Figure \ref{3fiegs} (as a side note, notice that the asymptotic behavior is reached faster for this smaller value of $\theta_{\text{MLE}}=22$).

However, it is important to stress that the Gaussian shape and consequently the approximation \eqref{mlelr} is not empirically supported for small databases of size $n=100$.

\begin{figure}[htbp]

\medskip
\hspace{0.35\baselineskip}\hfil
\makebox[0.4\textwidth]{(a) Comparisons}\hfil
\makebox[0.4\textwidth]{(b) Differences}\hfil

%\settoheight{\tempdim}{\includegraphics[width=0.3\textwidth]{test1_final.pdf}}%
%\rotatebox{90}{\makebox[\tempdim]{Test 1}}\hfil
\includegraphics[width=0.4\textwidth]{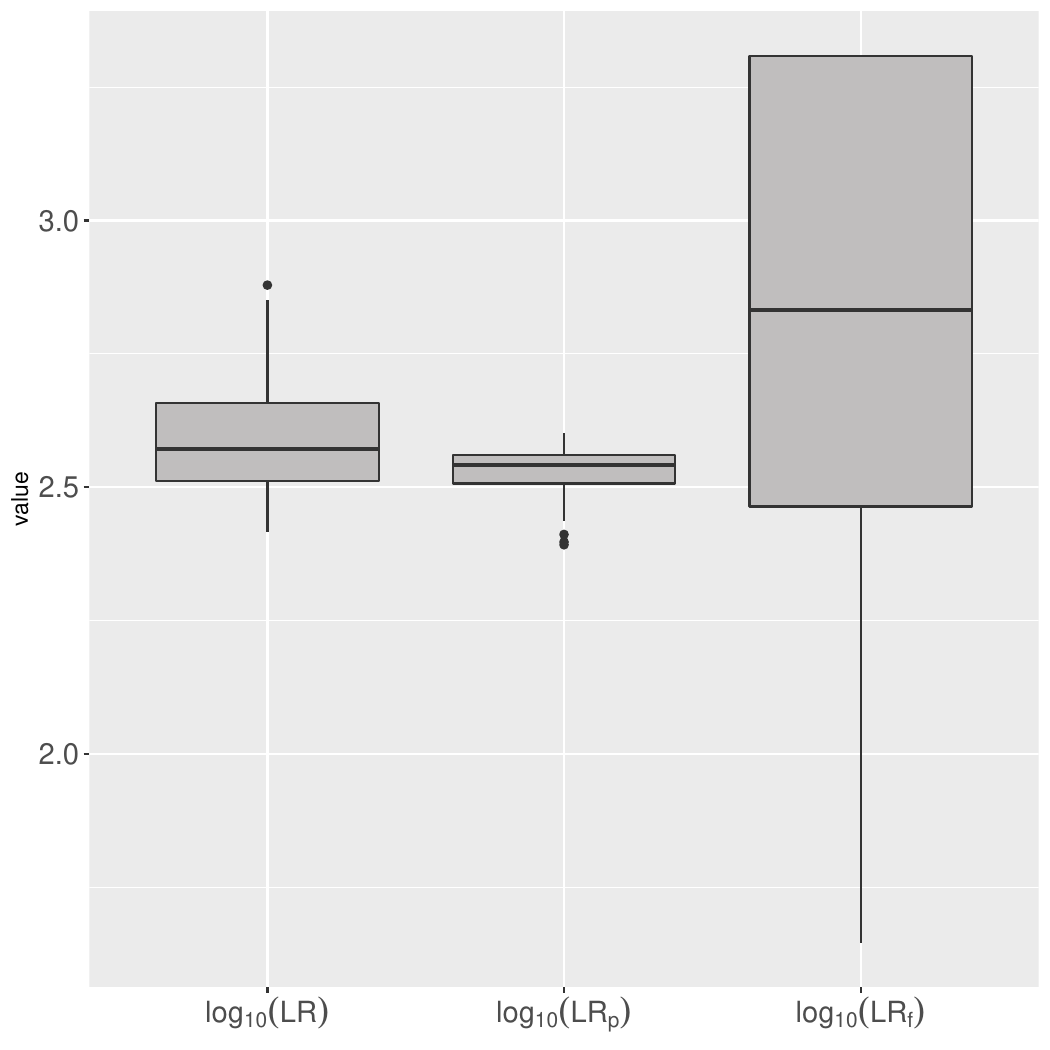}\hfil
\includegraphics[width=0.4\textwidth]{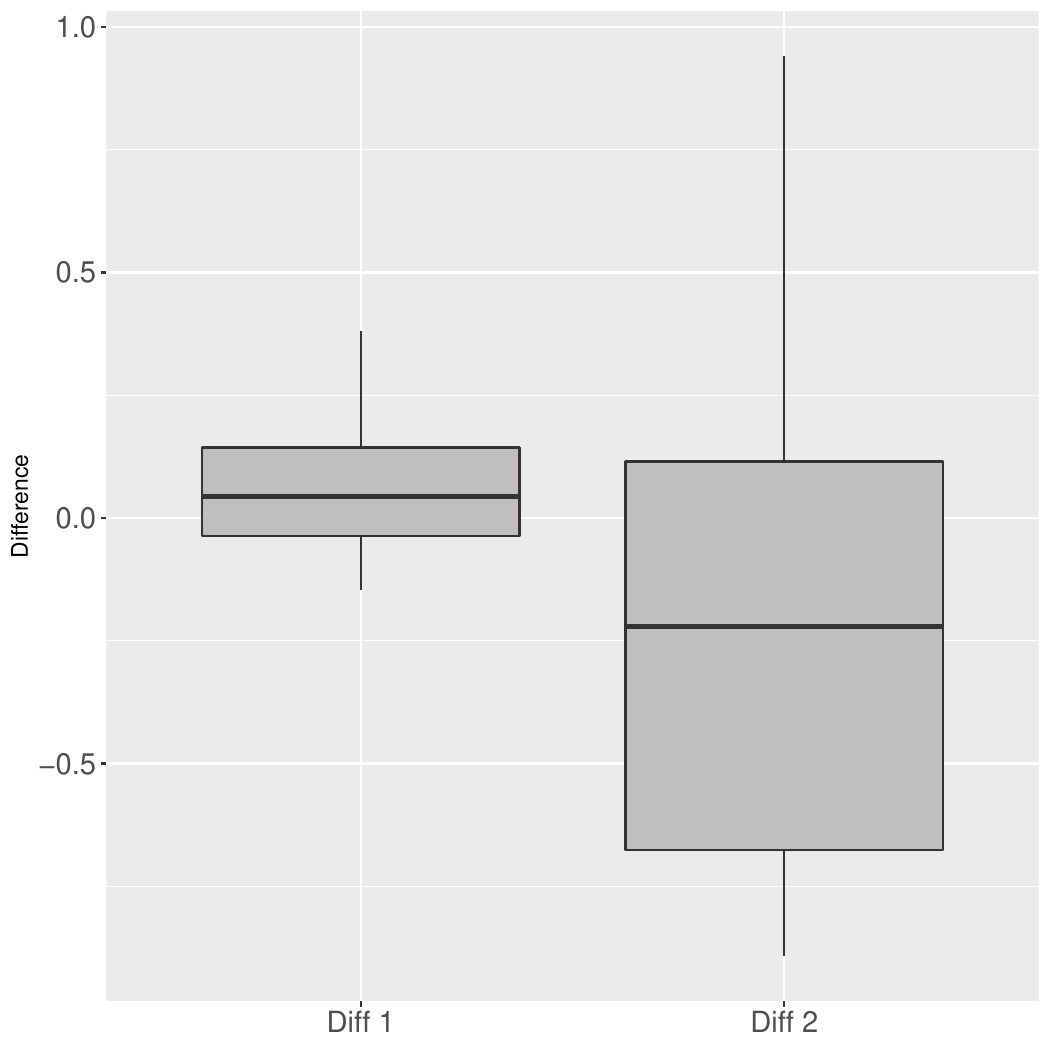}\hfil

\caption{\small (a) comparison between the distribution of $\log_{10}\mathrm{LR}$, $\log_{10}\mathrm{LR}_{|\mathbf{p}}$, and $\log_{10}\LR_{f}$. (b) the error $\log_{10}\LR - \log_{10}\LR_{\mid\mathbf{p}}$ and $\log_{10}\LR - \log_{10}\LR_{f}$. }\label{figggag3}
 
\end{figure}

In Table~\ref{tab1} and Figure~\ref{figggag3} (a) we compare the distribution of $\log_{10}\LR_{|\mathbf{p}}$, $\log_{10}\LR$, and $\log_{10}\LR_{f}$ obtained by 96 samples of size 100 from the Dutch population . Each sample represents a different rare type match case with a specific database of reference of size $n=100$.

The distribution of the benchmark likelihood ratio $\log_{10}\LR_f$ has more variation than the distribution of the Bayesian likelihood ratio, while $\log_{10}\LR_{|\mathbf{p}}$ appears to be the most concentrated around its mean. 
%This is probably due to the small size of the population and of the sampled databases.

In Table~\ref{tab2} and Figure~\ref{figggag3} (b) we consider the distribution of the two differences, Diff$_1$ and Diff$_2$.
Diff$_1$ is the smallest and the most concentrated: it ranges between -0.146 and 0.381 and has a small standard deviation. It means that the nonparametric Bayesian likelihood ratio obtained as in~\eqref{mlelr} can be thought of as a good approximation of the frequentist likelihood ratio for the same reduction of data ($\log_{10}\LR_{\mid\mathbf{p}}$), even though we have not empirically validated the approximation for small databases of size 100.
This difference is due to three things: the approximation \eqref{mlelr}, the MLE estimation of the hyperparameters, and the choice of a prior distribution (two-parameter Poisson Dirichlet) which is quite realistic, as shown in Figure~\ref{3fiegs}, but not perfectly fitting the actual population.
\begin{table}[htbp]
 
\centering
    \begin{tabular}{|l|c|c|c|c|c|c|c|}
    \hline
  &  Min   & 1st Qu. & Median & Mean  & 3rd Qu. & Max   & sd   \\
  \hline
 	 $\log_{10}\LR$   &   2.417 &  2.512 &  2.572 &  2.59&   2.658 &  2.879 	& 	0.102   \\
	$\log_{10}\LR_{|\mathbf{p}}$    	& 2.392  & 2.507   &2.542  & 2.529   &2.56&   2.602 &	0.045 \\
	$\log_{10}\LR_{f}$ &  1.646   &2.464  & 2.832 &  2.803 &  3.309 &3.309 	&	0.463 \\				
%$\widehat{\log_{10} LR}$ without   &  1.614 &  4.362 &  5.603  & 5.515 &  6.592 & 11.270    & 1.576 \\
   
   \hline
    \end{tabular}

    \caption{\small Summaries of the distribution of $\log_{10}\mathrm{LR}$, $\log_{10}(\mathrm{LR}_{|\mathbf{p}})$, and $\log_{10}\mathrm{LR}_{f}$. }\label{tab1}
 
\end{table}

\begin{table}[htbp]
\begin{center}
    \begin{tabular}{|l|c|c|c|c|c|c|c|}
    \hline
  &  Min   & 1st Qu. & Median & Mean  & 3rd Qu. & Max   & sd   \\
  \hline
 	 Diff$_1$   				&  -0.146 & -0.036 & 0.044  &0.06& 0.144  &0.381 &  0.126  \\
	 Diff$_2$       	&  -0.891 &-0.676 &-0.221& -0.213 & 0.115 & 0.94 & 	0.472  \\
				
%$\widehat{\log_{10} LR}$ without   &  1.614 &  4.362 &  5.603  & 5.515 &  6.592 & 11.270    & 1.576 \\
   
   \hline
    \end{tabular}

    \caption{\small Summaries of the distribution of $\mathrm{Diff}_1$, $\mathrm{Diff}_2$, and $\mathrm{Diff}_3$. }\label{tab2}
\end{center}
\end{table}

Notice that the difference increases if the Bayesian nonparametric likelihood is compared to the benchmark likelihood ratio (Diff$_2$). \textcolor{black}{Here, the reduction of data comes into play, too.} However, the difference ranges within one order of magnitude, but most of the time lies between -0.676 and 0.115, thus small.

\section{Conclusion}
This paper discusses the first application of a Bayesian nonparametric method to likelihood ratio assessment in forensic science, in particular to the challenging situation of the rare type match. If compared to traditional Bayesian methods such as those described in \citet{cereda:2015}, it presents many advantages. First of all, the prior chosen for the parameter $\mathbf{p}$ is more realistic for the population whose frequencies we want to model. Moreover, despite the theoretical background on which it lies may seem very technical and difficult, the method is extremely simple in practice, thanks to the use of an empirical Bayes approximation.
More could be done in the future: in particular regarding approximation \eqref{mlelr}. The posterior expectation in the denominator could, for instance, be treated using MCMC algorithms or ABC algorithms. Then, we can try to improve the efficiency of the Metropolis Hashting algorithm defined in Section \ref{tlr} in order to be used with bigger and better populations. The big problem is how to use these methods when relevant populations are poorly defined and accessible data-bases are of doubtful relevance. We don't solve those problems.

It is not clear whether other methods are better. This is all very open and controversial. We suggest the analyst to perform several very different analyses and think carefully what the differences between the conclusions tells her.
With this aim, we plan to compare this Bayesian nonparametric method to other existing methods for the rare type match problem, investigating calibration and validation through the use of ECE plots \citep{ramos:2013}.
\section*{Acknowledgment}
We are indebted to Jim Pitman and Alexander Gnedin for their help in understanding their important theoretical results, to Mikkel Meyer Andersen for providing a cleaned version of the database of \citet{purps:2014} and to Stephanie Van der Pas, Pierpaolo De Blasi and Giacomo Aletti for their useful opinions and comments.
This research was supported by the Swiss National Science Foundation, through grants no.  P2LAP2$\_$178195 

\bibliographystyle{apalike} 
%%%%%%%%%%%%%%%%%%%%%%%%%%%%%%%%%%%%%%%%%%
%=====================================

% The following MDPI journals use author-date citation: Arts, Econometrics, Economies, Genealogy, Humanities, IJFS, JRFM, Laws, Religions, Risks, Social Sciences. For those journals, please follow the formatting guidelines on http://www.mdpi.com/authors/references
% To cite two works by the same author: \citeauthor{ref-journal-1a} (\citeyear{ref-journal-1a}, \citeyear{ref-journal-1b}). This produces: Whittaker (1967, 1975)
% To cite two works by the same author with specific pages: \citeauthor{ref-journal-3a} (\citeyear{ref-journal-3a}, p. 328; \citeyear{ref-journal-3b}, p.475). This produces: Wong (1999, p. 328; 2000, p. 475)

%%%%%%%%%%%%%%%%%%%%%%%%%%%%%%%%%%%%%%%%%%
%% optional%% for journal Sci
%\reviewreports{\\
%Reviewer 1 comments and authors’ response\\
%Reviewer 2 comments and authors’ response\\
%Reviewer 3 comments and authors’ response
%}

%%%%%%%%%%%%%%%%%%%%%%%%%%%%%%%%%%%%%%%%%%
\end{document}